\documentclass[journal]{IEEEtran}

\usepackage{amssymb}
\usepackage{amsmath}
\usepackage{graphicx}
\usepackage{tcolorbox}

\usepackage[ruled]{algorithm2e}
\usepackage{bm}
\usepackage{romannum}
\usepackage{stfloats}
\usepackage{tabularx,ragged2e,booktabs,caption}
\usepackage{xcolor}
\usepackage{enumitem}
\usepackage{hyperref}
\hypersetup{
    colorlinks=true,
    linkcolor=black,
    filecolor=black,      
    urlcolor=black,
}
\newcommand{\norm}[1]{\left\lVert#1\right\rVert}
\newcommand{\note}[1]{{\color{black} {#1}}}
\DeclareMathOperator*{\argmax}{arg\,max}
\DeclareMathOperator*{\argmin}{arg\,min}

\usepackage{siunitx}
\begin{document}
\title{Learning to Demodulate from Few Pilots  via \\ Offline and Online Meta-Learning}
\author{Sangwoo Park,~\IEEEmembership{Student Member,~IEEE,}
        Hyeryung Jang,~\IEEEmembership{Member,~IEEE,}\\
        Osvaldo Simeone,~\IEEEmembership{Fellow,~IEEE,}
        and~Joonhyuk Kang,~\IEEEmembership{Member,~IEEE}

\thanks{This work was presented in part at IEEE SPAWC 2019 \cite{park2019learning}.} 
\thanks{S. Park and J. Kang are with the Department of Electrical Engineering, Korea Advanced Institute of Science and Technology, Daejeon 34141, South Korea (e-mails: \{sangwoop, jkang\}@kaist.ac.kr).}   
\thanks{H. Jang and O. Simeone are with the Department of Informatics, King's College London, London WC2R 2LS, U.K. (e-mails: \{hyeryung.jang, osvaldo.simeone\}@kcl.ac.uk).}
\thanks{\textcolor{black}{The work of S. Park was supported by Institute of Information $\&$ Communications Technology Planning $\&$ Evaluation (IITP) grant funded by the Korea Government (MSIT) (No.2018-0-00170, Virtual Presence in Moving Objects through 5G).}}
\thanks{The work of H. Jang and O. Simeone was supported by the European Research Council (ERC) under the European Union's Horizon 2020 research and innovation programme (grant agreement No. 725731).}
\thanks{\textcolor{black}{The work of J. Kang work was supported in by the Ministry of Science and ICT (MSIT), South Korea, through the Information
Technology Research Center (ITRC) Support Program supervised by the Institute of Information and Communications Technology
Planning and Evaluation (IITP) under Grant IITP-2020-0-01787.}}
}


\pagenumbering{arabic}

\maketitle
\thispagestyle{plain}
\pagestyle{plain}

\begin{abstract}
This paper considers an Internet-of-Things (IoT) scenario in which devices \textcolor{black}{sporadically transmit} short packets with few pilot symbols over a fading channel. Devices are characterized by unique transmission non-idealities, such as \note{I/Q imbalance}. The number of pilots is generally insufficient to obtain an accurate estimate of the end-to-end channel, which includes the effects of fading and of the transmission-side distortion. This paper proposes to tackle this problem by using meta-learning. Accordingly, pilots from previous IoT transmissions are used as meta-training data in order to train a demodulator that is able to quickly adapt to new end-to-end channel conditions from few pilots. Various state-of-the-art meta-learning schemes are adapted to the problem at hand and evaluated, including \note{Model-Agnostic Meta-Learning} (MAML), \note{First-Order MAML} (FOMAML), REPTILE, and \note{fast  Context  Adaptation  VIA  meta-learning} (CAVIA). Both offline and online solutions are developed. In the latter case, an integrated online meta-learning and adaptive pilot number selection scheme is proposed. Numerical results validate the advantages of meta-learning as compared to training schemes that either do not leverage prior transmissions or apply a standard joint learning algorithms on previously received data.
\end{abstract}



%
\IEEEpeerreviewmaketitle

\vspace{0.2cm}
\section{Introduction}
\label{sec:intro}

\subsection{Motivation}
\label{subsec:motivation}

For many standard channel models, such as additive Gaussian noise and fading channels with receive Channel State Information (CSI), the design of optimal demodulators and decoders is well understood. Most communication links hence use pilot sequences to estimate CSI, which is then plugged into the optimal receiver with ideal receive CSI (see, e.g., \cite{ostman2018short}). This standard model-based approach is inapplicable if: (\emph{i}) an accurate channel model is unavailable; and/or (\emph{ii}) the optimal receiver for the given transmission scheme and channel is of prohibitive complexity or unknown. Examples of both scenarios are reviewed in \cite{ibnkahla2000applications}, \cite{simeone2018very}, and include new communication set-ups, such as molecular channels, which lack well-established models; and links with strong non-linearities, such as satellite \note{channels} with non-linear transceivers, whose optimal demodulators can be highly complex \cite{ibnkahla2000applications}, \cite{bouchired1998equalisation}. This observation has motivated a long line of work on the application of machine learning methods to the design of demodulators or decoders, from the 90s \cite{ibnkahla2000applications} to many recent contributions, including \cite{o2017introduction, dorner2017deep, khani2019adaptive} and references therein.

Demodulation and decoding can be interpreted as classification tasks, \note{where} the input is given by the received baseband signals and the output consists of the transmitted symbols for demodulation, and of the transmitted binary messages for decoding. Pilot symbols can hence be used as training data to carry out the supervised learning of a parametric model for the demodulator or decoder, such as  Support Vector Machines (SVMs) or neural networks. The performance of the trained ``machine" as a demodulator or a decoder generally depends on how representative the training data \note{are} for the channel conditions encountered during test time and on the suitability of the parametric model in terms of trade-off between bias and variance \cite{hastie2009elements}.

\begin{figure}
    \centering
    \includegraphics[width=0.74\columnwidth]{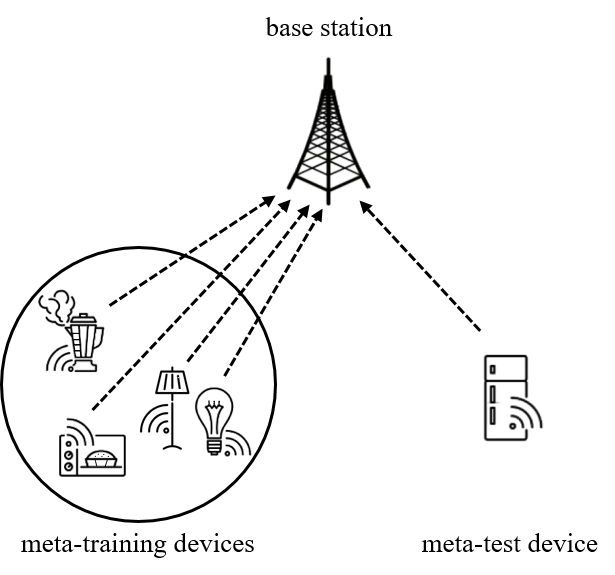}
    \caption{Illustration of few-pilot training for an IoT system via meta-learning.}
    \label{fig:iot_system}
\end{figure}

To the best of our knowledge, all the prior works reviewed above assume that training is carried out using pilot signals from the same transmitter whose data is to be demodulated or decoded. This generally requires the transmission of long pilot sequences for training. In this paper, we consider an Internet-of-Things (IoT)-like scenario, illustrated in Fig.~\ref{fig:iot_system}, in which devices \textcolor{black}{sporadically transmit} short packets with few pilot symbols. The number of pilots is generally insufficient to obtain an accurate estimate of the end-to-end channel, which generally includes the effects of fading and of the transmitter's non-linearities \cite{helmy2017robustness}. We propose to tackle this problem by using \emph{meta-learning} \cite{thrun1998lifelong}. 

\subsection{Meta-Learning}
\label{subsec:meta-learning}
Meta-learning, also sometimes referred to as ``learning to learn", aims at leveraging training and test data from different, but related, tasks for the purpose of acquiring an inductive bias that is suitable for the entire class of tasks of interest \cite{thrun1998lifelong}. The inductive bias can \note{be optimized by selecting either a model class, e.g., through a feature extractor, or a training algorithm, e.g., through} an initialization of model parameters or \note{the learning rate} \cite{grant2018recasting, simeone2020learning}. An important application of meta-learning is the acquisition of a \note{training procedure} that allow\note{s} a quick adaptation to a new, but related, task using few training examples, also known as \emph{few-shot learning} \cite{vinyals2016matching}. For instance, one may have training and test labelled images for binary classifiers of different types of objects, such as cats vs dogs or birds vs bikes\note{. These} can be used as meta-training data to quickly learn a new binary classifier, say for handwritten \textcolor{black}{binary digits}, from few training examples. 

Meta-learning has recently received renewed attention, particularly thanks to advances in the development of methods based on Stochastic Gradient Descent (SGD), including Model-Agnostic Meta-Learning (MAML) \cite{finn2017model}, REPTILE \cite{nichol2018first}, and fast Context Adaptation VIA meta-learning (CAVIA) \cite{zintgraf2019fast}. Such techniques can be generally classified as either \emph{offline}, \note{in which case} the meta-training data is fixed and given \cite{finn2017model, nichol2018first, zintgraf2019fast}; or \emph{online}, \note{in which case} all prior data from related tasks is treated as meta-training data in a streaming fashion \cite{finn2019online}.

\begin{figure}[t!]
    \centering
    \includegraphics[width=0.65\columnwidth]{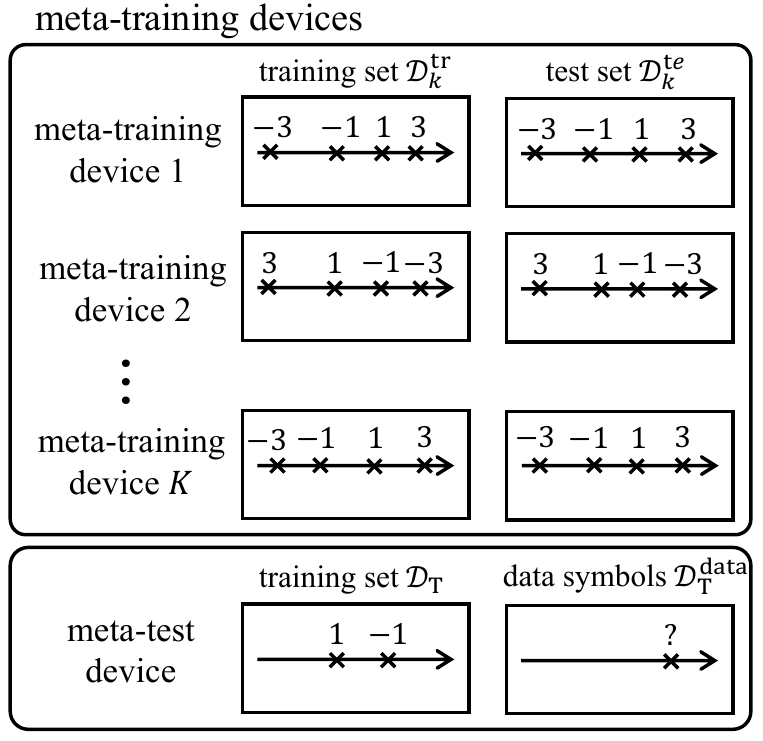}
    \caption{Offline meta-learning: Meta-training and meta-test data for $4$-PAM transmission from set $\mathcal{S} = \left\{-3,-1,1,3\right\}$. The figure assumes $N=8$ pilot symbols divided into $N^\text{tr}=4$ for meta-training and $N^\text{te}=4$ for meta-testing, and $P=2$ pilots for the meta-test device. Crosses represent received signals $y_k^{(n)}$, and the number above each cross represents the corresponding label, i.e., the pilot symbol $s_k^{(n)}$.}
    \label{fig:meta_4pam}
\end{figure}

\begin{figure}[t!]
    \centering
    \includegraphics[width=0.98\columnwidth]{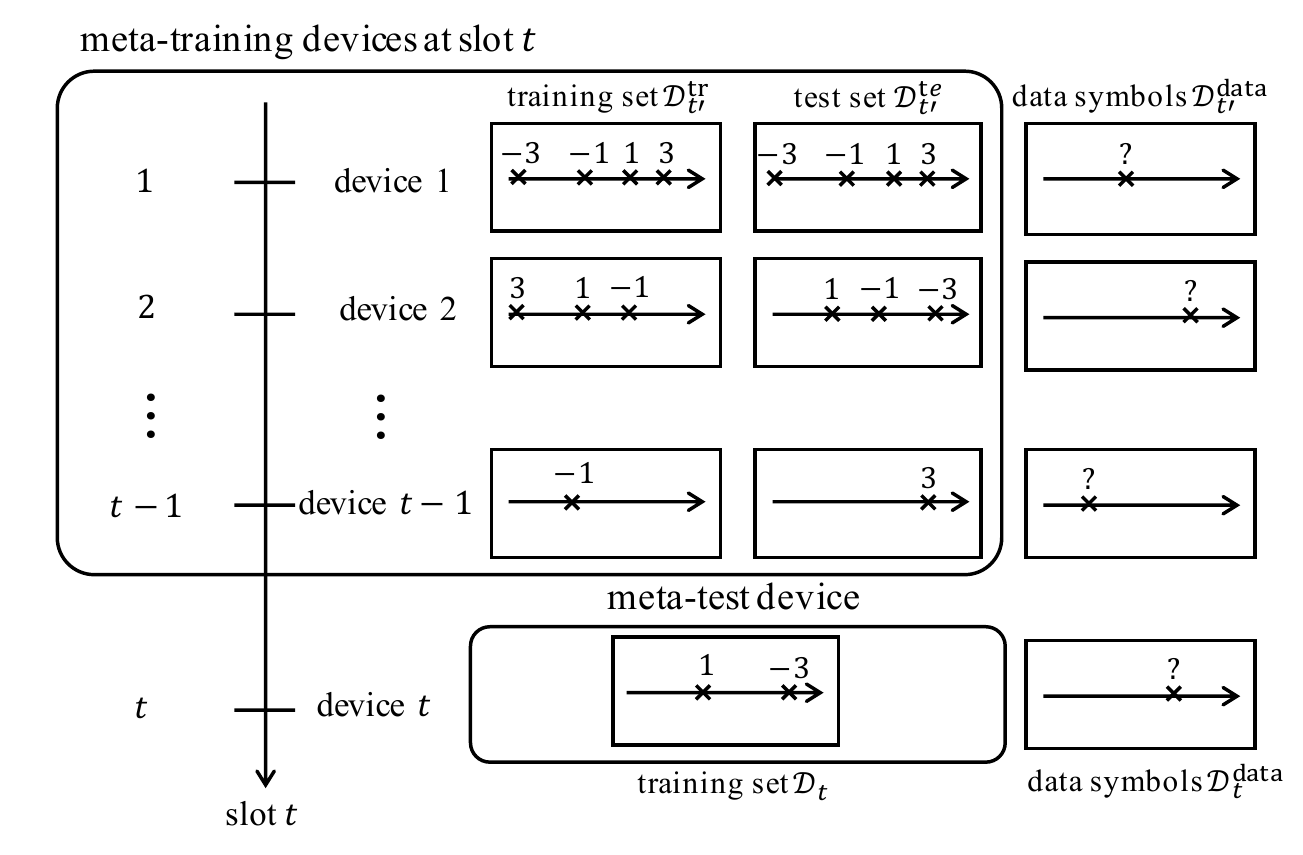}
    \caption{Online meta-learning: Meta-training and meta-test data for $4$-PAM transmission from set $\mathcal{S} = \left\{-3,-1,1,3\right\}$. Meta-training data are accumulated as the BS observes subsequent slots $t=1,2,\ldots$, with one device transmitting pilots and data symbols in each slot.}
    \label{fig:online-meta_4pam}
\end{figure}

\subsection{Main Contributions}
\label{subsec:main-contributions}
As illustrated in Fig\note{s}.~\ref{fig:meta_4pam} and \ref{fig:online-meta_4pam}, the key idea of this paper is to use pilots from previous transmissions of other IoT devices as meta-training data in order to train a procedure that is able to quickly adapt a demodulator to new end-to-end channel conditions from few pilots. We consider both an offline formulation, \note{in which} the set of previous transmissions is fixed, and an online set-up, in which the meta-training set is updated as transmitted pilots are received. The main contributions are as follows:
\begin{itemize}
    \item We adapt to the problem at hand a number of state-of-the-art offline meta-learning solutions, namely MAML \cite{finn2017model}, \note{First-Order MAML (FOMAML)} \cite{finn2017model}, REPTILE \cite{nichol2018first}, and CAVIA \cite{zintgraf2019fast}. 
    \textcolor{black}{Their relative merits and a unified interpretation in terms of the Expectation-Maximization (EM) algorithm are discussed;} 
    \item We validate the advantage of meta-learning with extensive numerical results \note{that provide comparisons with conventional model-based and learning-based communication schemes}. 
    \textcolor{black}{A comparative study of the performance of various meta-learning solutions is also presented;} 
    \item We propose a novel online solution that integrates meta-learning with an adaptive selection of the number of pilots and compare the proposed solution with conventional non-adaptive solutions in terms of receiver's performance and number of pilots.
\end{itemize}

The results in this paper have been partially presented in \cite{park2019learning}. In particular, reference \cite{park2019learning} derives an offline MAML-based algorithm, and offers some preliminary numerical results. 
\note{As compared to the preliminary conference version [1], this paper presents additional analysis, including a general framework for meta-learning based on EM; novel algorithms, introducing a comprehensive evaluation of a larger number of meta-learning offline schemes and the study of online meta-learning for demodulation, along with a new adaptive pilot \textcolor{black}{number selection} algorithm; and more extensive discussions \textcolor{black}{in terms of both} algorithm definition and \textcolor{black}{extra} experiments.} 

\subsection{Related Works}
\label{subsec:related_works}
In \cite{jiang2019mind}, which is concurrent to \cite{park2019learning}, the authors train a neural network-based decoder that can adapt to the new channel condition with a minimal number of pilot symbols using meta-learning via FOMAML. In \cite{mao2019metaicc}, the authors train a neural network-based channel estimator in OFDM system with meta-learning via FOMAML in order to obtain an effective channel estimation given a small number of pilots. \note{After the first submission of this paper, several additional papers have considered meta-learning for communication. Reference \cite{simeone2020learning} provides a review of meta-learning with applications to communication systems. In \cite{yang2019deep}, meta-learning is used  for downlink/uplink channel conversion in Frequency-Division Duplex massive MIMO channels. Papers \cite{park2019meta} and \cite{park2020end} consider meta-learning for end-to-end training of physical layer with and without a channel model, respectively. Finally, reference \cite{goutay2020deep} considers a related approach based on hypernetworks to aid neural network-based MIMO detection.}

\note{\emph{Paper organization}:} The rest of the paper is organized as follows. In Sec.~\ref{sec:model}\note{,} we detail system model and offline meta-learning problem. \textcolor{black}{In Sec.~\ref{sec:meta}\note{,} various offline meta-learning solutions are covered within a unified interpretation.} In Sec.~\ref{sec:online}\note{,} we redefine system model for an online setting and propose a novel online solution, including adaptive pilot \textcolor{black}{number selection}. Numerical results are presented in Sec.~\ref{sec:exp} and \note{conclusions are presented} in Sec.~\ref{sec:conclusions-future-works}.   
\vspace{0.2cm}
\section{Model and Problem}
\label{sec:model}

\subsection{System Model}

In this paper, we consider the IoT system illustrated in Fig.~\ref{fig:iot_system}, which consists of a number of devices and a base station (BS). For each device $k$, \textcolor{black}{the complex symbol transmitted by the device and the corresponding received signal at the BS are denoted as $s_k \in \mathcal{S}$ and $y_k$, respectively.} 
We also denote \textcolor{black}{as} $\mathcal{S}$ the set of all constellation symbols as determined by the modulation scheme. The end-to-end channel for a device $k$ is defined as 
\begin{align} \label{eq:e2e-channel}
y_k = h_k x_k + z_k,
\end{align}
where $h_k$ is the complex channel gain from device $k$ to the BS, \note{which is constant over a transmission block according to the standard quasi-static fading model typically assumed for short-packet transmissions};
$z_k \sim \mathcal{C}\mathcal{N}(0,N_0)$ is additive white complex Gaussian noise; and 
\begin{align} \label{eq:hardware_distortion_rand}
x_k \sim p_k(\cdot|s_k)
\end{align}
is the output of a generally random transformation defined by the conditional distribution $p_k(\cdot|s_k)$. This conditional distribution accounts for transmitter's non-idealities such as phase noise \cite{costa2002m}, I/Q imbalance \cite{windisch2005performance}, and amplifier's characteristics \cite{helmy2017robustness} of the IoT device. \textcolor{black}{Throughout the paper for each device $k$, we assume pilots and data symbols to follow the same constellation $S$ and to be subject to the transmitter's non-idealities defined by $p_k(\cdot|s_k)$.} The \textcolor{black}{average transmitted} energy per symbol is constrained as $\mathbb{E}[|x_k|^2]\leq E_x$ for some positive value $E_x$ \textcolor{black}{for both pilot and data symbols.}
As an example for \textcolor{black}{the transmitter's non-idealities \eqref{eq:hardware_distortion_rand}}, a common model for the I/Q imbalance assumes the following transformation \cite{tandur2007joint}
\begin{align}\label{eq:iq_imbalance}
\nonumber
x_k = &(1+\epsilon_k)\cos \delta_k \text{Re}\{s_k\} - (1+\epsilon_k)\sin \delta_k \text{Im}\{s_k\} \\ +  &j((1-\epsilon_k)\cos \delta_k \text{Im}\{s_k\} - (1-\epsilon_k)\sin \delta_k \text{Re}\{s_k\}),
\end{align}
where $\epsilon_k$ and $\delta_k$ represent the amplitude imbalance factor and phase imbalance factor, which are real constants or random variables. Note that we only explicitly \textcolor{black}{model imperfections} at the transmitter side. This is because, in practice, non-idealities on the receiver processing chain at the BS are expected to be much less significant than the mentioned non-idealities for IoT devices. Furthermore, receiver-side non-linearities at the BS can be also mitigated through offline designs prior to deployment.

Based on the reception of \note{few} pilots from a target device, we aim at determining a demodulator that \note{correctly} recovers the transmitted symbol $s$ from the received signal $y$ with high probability. The demodulator is defined by a conditional probability distribution $p(s | y, \varphi)$, which depends on a trainable parameter vector $\varphi$.

\subsection{Offline Meta-Learning Problem} 

Following the nomenclature of meta-learning \cite{finn2017model}, \textcolor{black}{the target device is referred to as the {\em meta-test device}}. To enable few-pilot learning, we assume here that the BS can use the signals received from the previous pilot transmissions of $K$ other IoT devices, which are referred to as {\em meta-training devices} and their data as \emph{meta-training data}. Specifically, as illustrated in Fig.~\ref{fig:meta_4pam}, the BS has \note{$N$ pairs} of pilot $s_k$ and received signal $y_k$ for each meta-training device $k=1,\ldots,K$. The meta-training dataset is denoted as $\mathcal{D} = \{\mathcal{D}_k \}_{k=1,\ldots,K}$, where $\mathcal{D}_k = \{ (s_k^{(n)}, y_k^{(n)}): n=1,\ldots,N \}$, and $(s_k^{(n)}, y_k^{(n)})$ are the \note{$n$-th} pilot-received signal pairs for the $k$th meta-training device. This scenario is referred to as offline meta-learning since the meta-training dataset $\mathcal{D}$ is fixed and given. Online meta-training will be discussed in Sec.~\ref{sec:online}.

For the target, or the meta-test, device, the BS receives $P$ pilot symbols. We collect the $P$ pilots received from the target device in set $\mathcal{D}_\text{T} = \{ (s^{(n)},y^{(n)}): n=1,\ldots,P\}$. 
The demodulator can be trained using meta-training data $\mathcal{D}$ and the pilot symbols $\mathcal{D}_\text{T}$ from the meta-test device. 

Training requires the selection of a parametric model $p(s|y,\varphi)$ for the demodulator. The choice of the parametric model $p(s|y,\varphi)$ should account for the standard trade-off between capacity of the model and overfitting \cite{bishop2006pattern, simeone2018brief}. To fix the ideas, we will assume that the demodulator $p(s|y,\varphi)$ is given by a multi-layer neural network with $L$ layers, with a softmax non-linearity in the final, $L$th, layer. This can be written as
\begin{align} \label{eq:softmax-demod}
p(s|y,\varphi) = \frac{\exp \Big( [f_{\varphi^{(L-1)}} ( f_{\varphi^{(L-2)}} (\cdot \cdot f_{\varphi^{(1)}}(y) ) )]_s \Big) }{ \sum\limits_{s' \in \mathcal{S}} \exp \Big( [f_{\varphi^{(L-1)}} ( f_{\varphi^{(L-2)}} (\cdot \cdot f_{\varphi^{(1)}}(y) ) )]_{s'} \Big) },
\end{align}
where $f_{\varphi^{(l)}}(x) = \sigma (W^{(l)}x+b^{(l)})$ represents the non-linear activation function of the $l$th layer with parameter $\varphi^{(l)}=\{ W^{(l)}, b^{(l)}\}$, with $W^{(l)}$ and $b^{(l)}$ being the weight matrix and bias vector of appropriate size, respectively; $[\cdot]_s$ stands for the element regarding $s$; and $\varphi = \{\varphi^{(l)}\}_{l=1,\ldots,L-1}$ is the vector of parameters. The non-linear function $\sigma(\cdot)$ can be, e.g., a \note{Rectified Linear Unit (ReLU)} or a hyperbolic tangent function. The input $y$ in \note{\eqref{eq:softmax-demod}} can be represented as a two-dimensional vector comprising real and imaginary parts of the received signal. 
\vspace{0.2cm}
\section{Offline Meta-learning Algorithms} 
\label{sec:meta}

In this section, we adapt state-of-the-art offline meta-learning algorithms \note{to} design \note{the demodulator in} \eqref{eq:softmax-demod} given meta-training and meta-test data. As discussed in Sec.~\ref{sec:intro}, we view demodulation as a classification task. 
To set the notation, for any set $\mathcal{D}_{0}$ of pairs $(s,y)$ of transmitted symbol $s$ and received signal $y$, \textcolor{black}{the standard cross-entropy loss function is defined as a function of the demodulator parameter vector $\varphi$ as }
\begin{align} \label{eq:loss-CE}
L_{\mathcal{D}_{0}}(\varphi) = -\sum_{(s,y) \in \mathcal{D}_{0}} \log p (s | y, \varphi).
\end{align}

\subsection{Joint Training}
\label{subsection:joint-training}
As a benchmark, we start by considering a conventional approach that uses the meta-training data $\mathcal{D}$ and the training data $\mathcal{D}_\text{T}$ for the \emph{joint training} of the model $p(s|y,\varphi)$. Joint training pools together all the pilots received from the meta-training devices and the meta-test device, and carries out the optimization of the cumulative loss $L_{\mathcal{D}\cup\mathcal{D_\text{T}}}(\varphi)$ in \eqref{eq:loss-CE} using SGD. Accordingly, the parameter vector $\varphi$ is updated iteratively based on the rule 
\begin{align} \label{eq:sgd}
\varphi \leftarrow \varphi + \eta \nabla_\varphi \log p(s^{(n)} | y^{(n)},\varphi),
\end{align}
by drawing one pair $(s^{(n)},y^{(n)})$ at random from the set $\mathcal{D}\cup\mathcal{D}_{\text{T}}$. In \eqref{eq:sgd}, the step size $\eta$ is assumed to be fixed for simplicity of notation but it can in practice be adapted across the updates (see, e.g., \cite{Goodfellow-et-al-2016}). Furthermore, this rule can be generalized by summing the gradient in \eqref{eq:sgd} over a mini-batch of pairs from the dataset $\mathcal{D}\cup\mathcal{D}_{\text{T}}$ at each iteration \cite{Goodfellow-et-al-2016}.

\begin{figure}
    \centering
    \includegraphics[width=0.45\columnwidth]{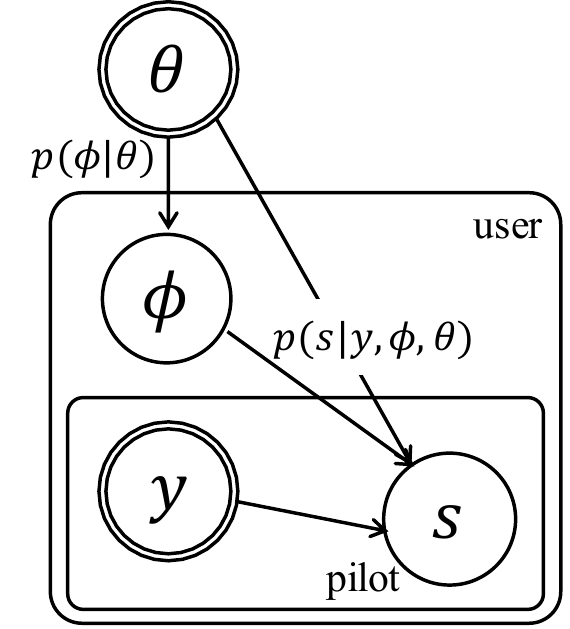}
    \caption{Graphical model assumed by meta-learning: The demodulator $p(s|y,\phi,\theta)$ depends on a user-specific, or context, random variable $\phi$, as well as on a shared parameter $\theta$, which may also affect the prior distribution of the context variable $\phi$. Double circles denote parameters, and the tile notation (see, e.g., \cite{koller2009probabilistic}) defines multiple users and pilots per user.}
    \label{fig:graphical-model}
\end{figure}

\subsection{A Unified View of Meta-Learning}
\label{subsection:unif-meta}
A useful way to introduce meta-learning in terms of the graphical model is illustrated in Fig.~\ref{fig:graphical-model}. Accordingly, meta-learning assumes a demodulator $p(s|y,\phi,\theta)$ that depends on a shared parameter $\theta$ common to all tasks, or users, and on a latent context variable $\phi$, which is specific to \note{each} user. The specific parameterization $p(s|y,\phi,\theta)$ and its relationship with \eqref{eq:softmax-demod} depend on the meta-learning scheme, and they will be discussed below. Note that, as illustrated in Fig.~\ref{fig:graphical-model}, the context vector $\phi$ is assumed to be random, while $\theta$ is a shared (deterministic) parameter. Furthermore, from Fig.~\ref{fig:graphical-model}, the shared variable $\theta$ can also affect the prior distribution of the context variable $\phi$. In this framework, the key idea is that meta-training data $\mathcal{D}$ \note{are} used to estimate the shared parameters $\theta$ via the process of meta-\note{learning}, while the context variable $\phi$ is inferred from the meta-test data $\mathcal{D}_\text{T}$. 

To elaborate, a principled way to train the model in Fig.~\ref{fig:graphical-model} would be to estimate parameter $\theta$ using the Expectation-Maximization (EM) algorithm based on the meta-training data $\mathcal{D}$. The EM algorithm is in fact the standard tool to tackle the problem of maximum likelihood estimation in the presence of latent variables, here the context \note{variable} $\phi$ (see, e.g., \cite{bishop2006pattern, simeone2018brief, koller2009probabilistic}). EM maximizes the sum of marginal likelihoods
\begin{align} \label{eq:em-training}
p(s|y,\theta)=\mathbb{E}_{\phi\sim p(\phi|\theta,\mathcal{D}_k)}[p(s|y,\phi,\theta)]
\end{align}
over the data pairs $(s,y)$ from all data sets $\mathcal{D}_k$ in the meta-training data set $\mathcal{D}$. In \eqref{eq:em-training}, the average is taken with respect to the posterior distribution $p(\phi|\theta,\mathcal{D}_k)$ of the context variable given the training data $\mathcal{D}_k$ of the $k$th meta-training device. 
After EM training, one can consider the obtained parameter $\theta$ as fixed when inferring a data symbol $s$ given a new observed signal $y$ and the pilots $\mathcal{D}_\text{T}$ for the meta-test device. This last step would ideally yield the demodulator 
\begin{align} \label{eq:em-estep}
p(s|y,\theta)=\mathbb{E}_{\phi\sim p(\phi|\theta,\mathcal{D}_\text{T})}[p(s|y,\phi,\theta)],
\end{align}
where the average is taken over the posterior distribution $p(\phi|\theta,\mathcal{D}_\text{T})$ of the context variable given the training data of the meta-test device. 

The computation of the posteriors $p(\phi|\theta,\mathcal{D}_k)$ in \eqref{eq:em-training} and $p(\phi|\theta,\mathcal{D}_\text{T})$ in \eqref{eq:em-estep} \note{is} generally of infeasible complexity. Therefore, state-of-the-art meta-learning techniques approximate this principled solution by either employing point estimate of latent context variable $\phi$ \cite{finn2017model, nichol2018first, zintgraf2019fast} or by \note{a} direct \note{approximation} of its posterior distribution \cite{ravi2018amortized, gordon2018meta, nguyen2019uncertaintybayes}. In this paper, we focus on the more common point estimate based meta-learning techniques, which are reviewed next.

\subsection{MAML}
\label{subsection:maml}
For any meta-training device $k$, MAML \cite{finn2017model} assumes a demodulator $p(s|y,\phi_k)$ given by \eqref{eq:softmax-demod} with model weights $\varphi$ equal to the context variable $\phi_k$. The user-specific variable $\phi_k$, rather than being obtained from the ideal posterior $p(\phi_k|\theta,\mathcal{D}_k)$ as in \eqref{eq:em-training}, is computed via SGD-based training from the data $\mathcal{D}_k$. Specifically, the key idea in MAML is to identify \textcolor{black}{an initial shared parameter $\theta$ during meta-training such that} starting from it, the SGD updates \eqref{eq:sgd} \note{for any meta-training device $k$ (i.e., for ${\mathcal{D}}_{0}= \mathcal{D}_k$)} produce a parameter vector $\phi_k$ that yields a low value of the loss function \eqref{eq:loss-CE}. As we will detail, it is possible to \note{include} one or multiple SGD updating steps \note{in} \eqref{eq:sgd} \cite{antoniou2018train}. After meta-training, the initial parameter $\theta$ is used for the SGD updates of the target device based on the pilots in set $\mathcal{D}_\text{T}$.

To elaborate, \textcolor{black}{assume first that the exact average loss $L_k(\phi_k)=\mathbb{E}[-\log p (s_k|\note{y_k},\phi_k)]$ for all meta-training devices $k=1,\ldots,K$ is given.} 
The average in $L_k(\phi_k)$ is taken over the distribution $p(s_k,\note{y_k}) = p(s_k)p(\note{y_k}|s_k)$, where $p(s_k)$ is the prior distribution of the transmitted symbol $s_k$ and $p(\note{y_k}|s_k)$ is defined by \eqref{eq:e2e-channel} \textcolor{black}{and \eqref{eq:hardware_distortion_rand}}. Note that in practice\note{,} this \note{distribution} is not \note{known} since the channel and the transmitters' model are not known a priori. During meta-training, MAML seeks an initial value $\theta$ such that, for every device $k$, the losses $L_k(\phi_k)$ obtained after one or more SGD updates starting from $\theta$ are collectively minimized. As discussed, the SGD updates can be interpreted as producing a point estimate of the context variables $\phi_k$ in the model in Fig.~\ref{fig:graphical-model} \cite{grant2018recasting}. Mathematically, with a single SGD iteration, \textcolor{black}{the estimate is obtained as}
\begin{align} \label{eq:maml-sgd-single}
\phi_k = \theta - \eta \nabla_\theta L_k(\theta).
\end{align}
More generally, with $m \geq 1$ local SGD updates we obtain $\phi_k = \phi^m_k$, where 
\begin{align} \label{eq:maml-sgd-multi}
\phi_k^{i}  &= \phi_k^{i-1} - \eta \nabla_{\phi^{i-1}_k}L_{k}(\phi^{i-1}_k),
\end{align}
for $i=1,\ldots,m,$ with $\phi^0_k = \theta$.
The identification of a shared parameter $\theta$ is done by minimizing the sum $\sum_{k=1}^K L_k(\phi_k)$ over $\theta$.

The losses $L_k(\phi_k)$ for all meta-training devices are not known and need to be estimated from the available data. To this end, in the meta-training phase, each set $\mathcal{D}_k$ of $N$ pairs of pilots and received signals for meta-training device $k$ is randomly divided into a training set $\mathcal{D}_k^{\text{tr}}$ of $N^\text{tr}$ pairs and a test set $\mathcal{D}_k^{\text{te}}$ of $N^\text{te}$ pairs, as shown in Fig.~\ref{fig:meta_4pam}. The updated context variable $\phi_k$ is computed by applying the SGD-based rule in \eqref{eq:sgd} \note{by using the} training subset $\mathcal{D}_k^\text{tr}$, as in \eqref{eq:maml-sgd-multi}, e.g., $\phi_k = \theta - \eta \nabla_\theta L_{\mathcal{D}_k^\text{tr}}(\theta)$ for a single update. \note{In practice, $m$ SGD updates can be carried out using mini-batches of training samples at each iteration.}
The loss $L_k(\phi_k)$ is then estimated by using the test subset $\mathcal{D}_k^\text{te}$ as $L_{\mathcal{D}_k^\text{te}}(\phi_k)$. Finally, MAML minimizes the overall estimated loss $\sum_{k=1}^K L_{\mathcal{D}_k^\text{te}}(\phi_k)$ by performing an SGD-based update in the \note{opposite} direction of the gradient $\nabla_\theta \sum_{k=1}^K L_{\mathcal{D}_k^\text{te}}(\phi_k)$ with step size $\kappa$. 

Considering first a single local SGD update \eqref{eq:maml-sgd-single} for the context variables, the meta-training update is finally given as 
\begin{align} \label{eq:maml-1}
\nonumber
\theta \leftarrow & \theta - \kappa \nabla_\theta \sum_{k=1}^K  L_{\mathcal{D}_k^{\text{te}}}(\phi_k) = \note{\theta - \kappa \sum_{k=1}^K} (\mathbf{J}_\theta \phi_k)\nabla_{\phi_k}L_{\mathcal{D}_k^{\text{te}}}(\phi_k)
\\=& \theta - \kappa \sum_{k=1}^K (I-\eta \nabla^2_{\theta}L_{\mathcal{D}_k^{\text{tr}}}(\theta))
\nabla_{\phi_k}L_{\mathcal{D}_k^{\text{te}}}(\phi_k),
\end{align}
where $\mathbf{J}_\theta$ represents the Jacobian operation, and $\kappa>0$ is \textcolor{black}{the} step size.
With multiple local SGD updating steps \note{in} \eqref{eq:maml-sgd-multi}, \textcolor{black}{the meta-training update can be similarly written as}  
\begin{align} \label{eq:maml-2}
\nonumber
\theta \leftarrow \theta &- \kappa \sum_{k=1}^K \Big[(I-\eta \nabla^2_{\theta}L_{\mathcal{D}_k^{\text{tr}}}(\theta))\cdots \\ 
&(I-\eta \nabla^2_{\phi^{m-1}_k}L_{\mathcal{D}_k^{\text{tr}}}(\phi^{m-1}_k))
\nabla_{\phi^m_k}L_{\mathcal{D}_k^{\text{te}}}(\phi^m_k)\Big].
\end{align}
\note{In practice, the meta-update \textcolor{black}{in} \eqref{eq:maml-1} \textcolor{black}{and} \eqref{eq:maml-2} can be carried out over a subset of meta-training devices at each iteration.}
Computation of the Hessian matrices needed in \eqref{eq:maml-1} and \eqref{eq:maml-2} can be significantly accelerated using a finite difference approximation for Hessian-vector product calculation \cite{lecun2012efficient}, \cite{andrei2009accelerated}, which is reviewed in Appendix~\ref{sec:appdendix-hess-vec-approx}. The MAML algorithm is summarized in Algorithm~\ref{alg:maml_fomaml_reptile}. 

\note{In Algorithm~\ref{alg:maml_fomaml_reptile}, meta-training is carried out for a given fixed number of iterations. Among these iterations, we choose the iterate that has the smallest meta-training loss $\sum_{k\in K'} L_{\mathcal{D}^\text{te}_k}(\phi_k)$, evaluated on the subset $K'$ of meta-training devices sampled for the corresponding meta-training update. Similarly, adaptation on the meta-test device chooses the iterate that has the lowest loss $L_{\mathcal{D}'_\text{T}}(\phi_\text{T})$, where $\mathcal{D}'_\text{T}$ is the mini-batch of training pairs $(s,y)$ drawn from $\mathcal{D}_\text{T}$ for the corresponding update.}

\subsection{FOMAML}
\label{subsectino:fomaml}
First-order MAML (FOMAML) \cite{finn2017model} is an approximation of MAML that ignores the second-derivative terms in the meta-training updates \eqref{eq:maml-1}--\eqref{eq:maml-2}. Accordingly, the meta-training update is given as
\begin{align} \label{eq:fomaml-1}
\theta \leftarrow \theta - \kappa \nabla_{\phi_k} \sum_{k=1}^K  L_{\mathcal{D}_k^{\text{te}}}(\phi_k).
\end{align}
As a result, FOMAML updates parameter $\theta$ in the \note{opposite} direction of the gradient $\nabla_{\phi_k} \sum_{k=1}^K  L_{\mathcal{D}_k^{\text{te}}}(\phi_k)$ instead of $\nabla_{\theta} \sum_{k=1}^K  L_{\mathcal{D}_k^{\text{te}}}(\phi_k)$.
For some neural network architectures and loss functions, e.g., networks with ReLU activation functions \cite{goodfellow2014explaining}, FOMAML has been reported to perform almost as well as MAML \cite{finn2017model}. \note{Algorithm~\ref{alg:maml_fomaml_reptile} provides a summary.}

\begin{algorithm}[h] 
\DontPrintSemicolon
\smallskip
\KwIn{Meta-training set $\mathcal{D} = \{\mathcal{D}_k\}_{k=1,\ldots,K}$ and pilots $\mathcal{D}_{\text{T}}$ from the target device; $N^\text{tr}$ and $N^\text{te}$; step size hyperparameters $\eta$ and $\kappa$; \note{number of meta-training iterations $I$; number of local updates during meta-training $m$ and meta-testing $I_\text{T}$}\\
\KwOut{Learned shared initial parameter vector $\theta$ and target-device specific parameter vector $\phi_\text{T}$}}
\vspace{0.15cm}
\hrule
\vspace{0.15cm}
{\bf initialize} parameter vector $\theta$ \\
\texttt{meta-learning phase} \\
\For{{\em $I$ meta-training iterations}}{
randomly select a subset $K'$ of meta-training devices\\
\For{{\em each \note{selected} meta-training device $k$}}{
randomly divide $\mathcal{D}_k$ into two sets $\mathcal{D}_k^{\text{tr}}$ of size $N^\text{tr}$ and $\mathcal{D}_k^{\text{te}}$ of size $N^\text{te}$ \\
do $m$ local updates for the context variable $\phi_k$ using \eqref{eq:maml-sgd-single}--\eqref{eq:maml-sgd-multi}
}
update shared parameter $\theta$ \note{via} \eqref{eq:maml-1}--\eqref{eq:maml-2} \note{for MAML}, \note{via} \eqref{eq:fomaml-1} \note{for FOMAML}, \note{via} \eqref{eq:reptile-1} \note{for REPTILE}}
return iterate $\theta$ with minimum meta-training loss $\sum_{k\in K'}L_{\mathcal{D}^\text{te}_k}(\theta)$\\
\vspace{0.2cm}
\texttt{adaptation on the meta-test device} \\
{\bf initialize} context parameter vector $\phi_{\text{T}} \leftarrow \theta$ \\
\For{{\em $I_\text{T}$ meta-testing iterations}}{
draw a \note{mini-batch $\mathcal{D}'_\text{T}$ of} pair\note{s} $(s^{(n)},y^{(n)})$ from $\mathcal{D}_\text{T}$ \\
update context variable $\phi_\text{T}$ in the direction of the gradient $\note{\sum_{(s,y)\in\mathcal{D}'_\text{T}}}\nabla_{\phi_\text{T}} \log p(\note{s| y},\phi_\text{T})$ with step size $\eta$ 
}
return iterate $\phi_\text{T}$ with minimum training loss $L_{\mathcal{D}'_\text{T} }(\phi_\text{T})$
\caption{Few-Pilot Demodulator Meta-Learning via MAML, FOMAML, REPTILE}
\label{alg:maml_fomaml_reptile}
\end{algorithm}

\subsection{REPTILE}
\label{subsection:reptile}
REPTILE \cite{nichol2018first} is a first-order gradient-based meta-learning algorithm as FOMAML. It uses the same local update \eqref{eq:maml-sgd-single}--\eqref{eq:maml-sgd-multi} for the context variables $\phi_k$, but the meta-training update is given as \note{
\begin{align} \label{eq:reptile-1}
\theta \leftarrow  (1-\kappa)\theta + \kappa \sum_{k=1}^K \phi_k.
\end{align}
The REPTILE update rule \eqref{eq:reptile-1} is essentially equivalent to that used for federated learning \cite{kairouz2019advances}.} We refer to \cite{nichol2018first} for a justification of the method.

\subsection{CAVIA}
\label{subsection:cavia}
Unlike the meta-learning techniques discussed so far, CAVIA \cite{zintgraf2019fast} interprets context variable $\phi$ as an additional input to the demodulator, so that the demodulator $p(s|y,\phi,\theta)$ can be written as in \eqref{eq:softmax-demod} with input given by the concatenation $\tilde{y}=[y, \phi]$ and model weights $\varphi$ equal to the shared parameter vector $\theta$. Using \eqref{eq:softmax-demod}, the demodulator is hence in the form $p(s|\tilde{y},\theta)$, where the shared parameter $\theta$ defines the weights of the demodulator model. After meta-training, the shared parameter $\theta$ is fixed, and the pilots in set $\mathcal{D}_\text{T}$ of the meta-test device are used to optimize the additional input vector $\phi$. 

\begin{algorithm}[h] 
\DontPrintSemicolon
\smallskip
\KwIn{Meta-training set $\mathcal{D} = \{\mathcal{D}_k\}_{k=1,\ldots,K}$ and pilots $\mathcal{D}_{\text{T}}$ from the target device; $N^\text{tr}$ and $N^\text{te}$; step size hyperparameters $\eta$ and $\kappa$; \note{number of meta-training iterations $I$; number of local updates during meta-training $m$ and meta-testing $I_\text{T}$}}
\KwOut{Learned parameter vector $\theta$ and target-device context parameter vector $\phi_\text{T}$}
\vspace{0.15cm}
\hrule
\vspace{0.15cm}
{\bf initialize} parameter vector $\theta$ \\
\texttt{meta-learning phase} \\
\For{{\em \note{$I$ meta-training iterations}}}{
randomly select a subset $K'$ of meta-training devices \\
\For{{\em each \note{selected} meta-training device $k$}}{
{\bf initialize} context parameter vector $\phi_k$ \\
randomly divide $\mathcal{D}_k$ into two sets $\mathcal{D}_k^{\text{tr}}$ of size $N^\text{tr}$ and $\mathcal{D}_k^{\text{te}}$ of size $N^\text{te}$ \\
do $m$ local updates for the context variable $\phi_k$ using \eqref{eq:cav-local}
}
update shared parameter $\theta$ \note{via} \eqref{eq:cav-meta-update}
}
\vspace{0.08cm}
return iterate $\theta$ with minimum meta-training loss $\sum_{k\in K'} L_{\mathcal{D}^\text{te}_k }(\theta)$\\
\vspace{0.48cm}
\texttt{adaptation on the meta-test device} \\
{\bf initialize} context parameter vector $\phi_{\text{T}}$ \\
\For{{\em \note{$I_\text{T}$ meta-testing iterations}}}{
draw a \note{mini-batch $\mathcal{D}'_\text{T}$ of} pair\note{s} $(s^{(n)},y^{(n)})$ from $\mathcal{D}_\text{T}$ \\
update context parameter vector $\phi_\text{T}$ in the direction of the gradient $\note{\sum_{(s,y)\in \mathcal{D}'_\text{T}}}\nabla_{\phi_\text{T}} \log p(\note{s | \tilde{y}},\theta)$  
with step size $\eta$ where \note{$\tilde{y} = [y,\phi_\text{T}]$}
}
\note{return iterate $\phi_\text{T}$ with minimum training loss $L_{\mathcal{D}'_\text{T} }(\theta)$}
\caption{Few-Pilot Demodulator Meta-Learning via CAVIA}
\label{alg:cavia}
\end{algorithm}

\note{During} meta-training, given the current value of the shared parameter $\theta$, the context variable $\phi_k$ is optimized by one or more SGD-based update\note{s} to minimize the loss $L_{\mathcal{D}^\text{tr}_k}(\theta)$ as
\begin{align} \label{eq:cav-local}
\phi_k \leftarrow &\phi_k - \eta \nabla_{\phi_k}  L_{\mathcal{D}^{\text{tr}}_k}(\theta).
\end{align}
Note that the loss $L_{\mathcal{D}^\text{tr}_k}(\theta)$ is a function of $\phi_k$ through the additional input $\phi_k$.
With the obtained additional input $\phi_k$, the meta-training update is given as
\begin{align} \label{eq:cav-meta-update}
\theta \leftarrow \theta - \kappa \nabla_\theta \sum_{k=1}^K  L_{\mathcal{D}_k^{\text{te}}}(\theta).
\end{align}
After meta-training, as mentioned, parameter $\theta$ is fixed, and the context vector $\phi_\text{T}$ is obtained by using SGD updates as 
\begin{align} \label{eq:cav-meta-test}
\phi_\text{T} \leftarrow &\phi_\text{T} - \eta \nabla_{\phi_\text{T}}  L_{\mathcal{D}_{\text{T}}}(\theta).
\end{align}
CAVIA is summarized in Algorithm~\ref{alg:cavia}.

\vspace{0.2cm}
\section{Online meta-learning Algorithm}
\label{sec:online}

\begin{figure}
    \centering
    \includegraphics[width=1\columnwidth]{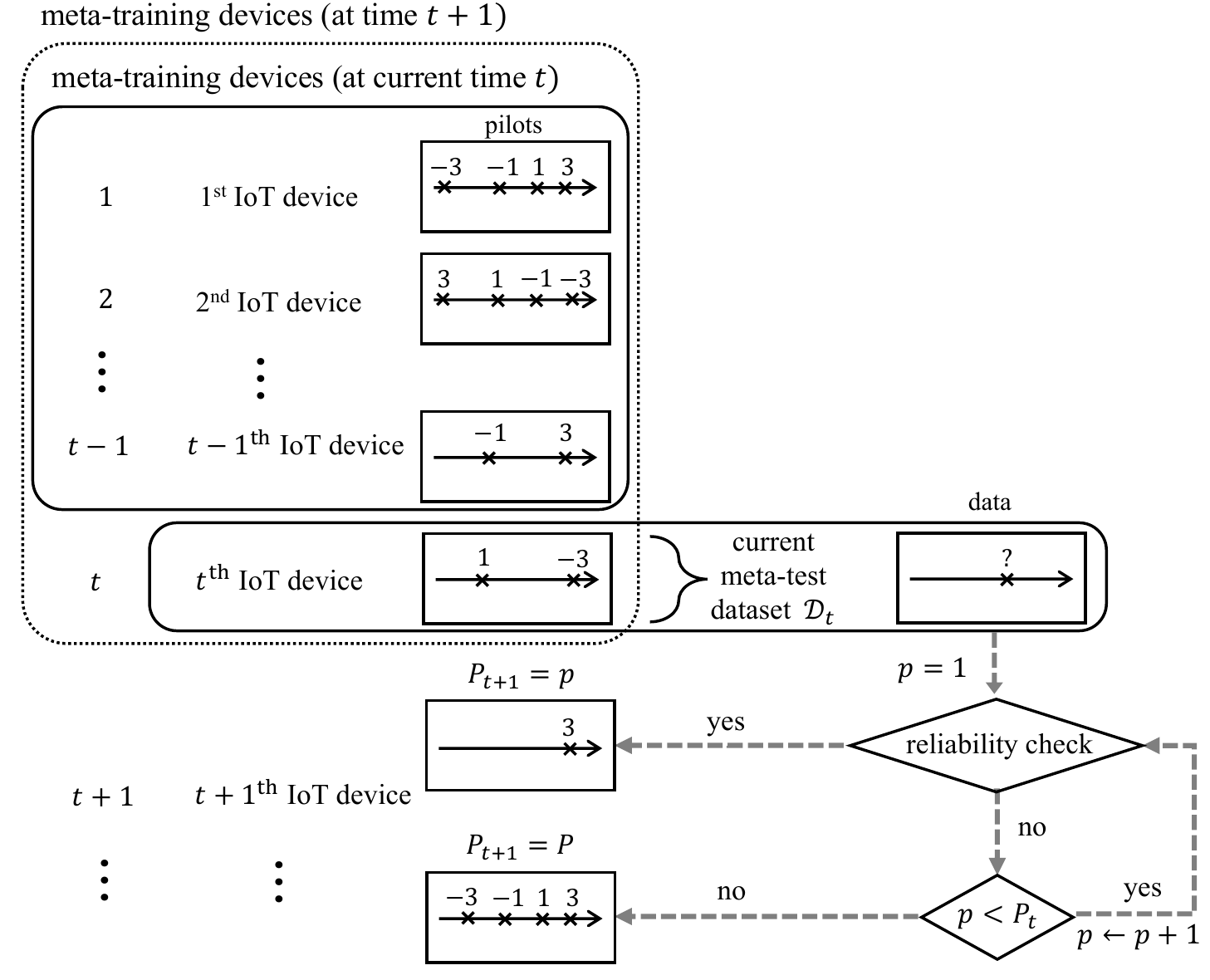}
    \caption{Illustration of adaptive pilot number selection based on online meta-learning for 4-PAM transmission. The number of transmitted pilots $P_{t+1}$ for the IoT device active in the next time slot is determined by the BS from the performance of the meta-learned demodulator in the current slot $t$.}
    \label{fig:online-diag}
\end{figure}

In this section, we consider an online formulation in which packets from devices, containing both pilots and a data payload, are sequentially received at the BS. Therefore, as illustrated in Fig.~\ref{fig:online-meta_4pam}, meta-training data are accumulated at the BS over time. The formulation follows the basic framework of online meta-learning introduced in \cite{finn2019online}, which proposes an online version of MAML. Here, we adapt the online meta-learning framework to the demodulation problem at hand, and extend it to integrate all the meta-training solutions discussed in the previous section, namely MAML, FOMAML, REPTILE, and CAVIA. 
\textcolor{black}{Moreover, a novel adaptive pilot number selection scheme is proposed that leverages the fast adaptation property of meta-learning to reduce the pilot overhead.} 

\subsection{System Model}
\label{subsection:online-sys-model}
As illustrated in Fig.~\ref{fig:online-meta_4pam}, in each slot $t=1,2,\ldots,$ the BS receives a packet from a new device, from which the BS obtains the set $\mathcal{D}_t=\{ (s^{(n)}_t,y^{(n)}_t):n=1,\ldots,N_t \}$ of $N_t$ pilots $\{ s^{(n)}_t \}$ and corresponding received signals $\{ y^{(n)}_t \}$. Each received packet also contains a payload of data $\mathcal{D}^{\text{data}}_t = \{ (y^{(n)}_t) :n=1,2,\ldots$\}. Therefore, at each slot $t$, the BS \note{has} meta-training data-carrying symbols $\mathcal{D}^{t-1}=\{ \mathcal{D}_{t'}\}^{t-1}_{t'=1}$ from previously active devices, as well as meta-test data $\mathcal{D}_t$ from the currently active device. The goal is training a demodulator $p(s|y,\varphi_t)$ that performs well on the payload data $\mathcal{D}^\text{data}_t$ after adaptation on the received pilots in set $\mathcal{D}_t$ by \note{using the} meta-training data $\mathcal{D}^{t-1}$ \note{as well}.

\subsection{Online Learning (Joint Training)}
\label{subsection:online-learning}
Before discussing online meta-learning, here we briefly summarize the standard online learning set-up as applied to the problem introduced above. \note{This} can be considered as the counterpart of joint training for the offline problem studied in Sec.~\ref{subsection:joint-training}. In online learning, the goal of the online learner is to determine a model parameter vector $\varphi_t$, sequentially at each slot $t$, that perform\note{s} well on the loss sequence $ L_{\mathcal{D}_t}(\varphi_t)$ for $t=1,2,\ldots$ (recall \eqref{eq:loss-CE}). As a benchmark, typical online learning formulations use the best single model $\varphi$ that can be obtained using knowledge of the losses $L_{\mathcal{D}_t}(\cdot)$ in hindsight for all relevant values of $t$, i.e., $\varphi\in \argmin_{\varphi} \sum_t L_{\mathcal{D}_t}(\varphi)$, where the sum is over the time horizon of interest \cite{shalev2012online}.

A standard online learning algorithm is Follow The Leader (FTL) \cite{hannan1957approximation}, which determines the parameter $\varphi_t$ that performs best on the previous data $\mathcal{D}^{t-1}$. For the problem at hand, FTL determines the parameter $\varphi_t$ at slot $t$ by tackling the problem 
\begin{align} \label{eq:ftl}
\varphi_{t} = \argmin_\varphi \sum_{k=1}^{t} L_{\mathcal{D}_k}(\varphi).
\end{align}
Note that in standard online learning formulations the sum in \eqref{eq:ftl} would be performed up to time \note{slot} $t-1$ due to the typical assumption that no data is a priori known at time $t$ about loss $L_{\mathcal{D}_t}(\cdot)$ \cite{shalev2012online}.
From \eqref{eq:ftl}, FTL can be interpreted as a form of joint training carried out in an online manner. From a theoretical standpoint, FTL can be shown to obtain a sub-linearly growing regret with respect to slot $t$ as compared to the discussed benchmark learner with hindsight information (see \cite{shalev2012online} for precise statements). 

\subsection{Online Meta-Learning}
\label{subsection:online-meta}
With meta-learning, as discussed in Sec.~\ref{subsection:unif-meta} (see Fig.~\ref{fig:graphical-model}), the demodulator $p(s|y,\phi,\theta)$ is defined by a shared parameter $\theta$ and by a context, device-dependent, variables $\phi$. \note{In the online setting}, in each slot $t$, we propose to estimate the shared parameter $\theta_t$ from the meta-training data $\mathcal{D}^{t-1}$, while the context variable $\phi_t$ for the currently active device is estimated from $\mathcal{D}_t$. 
These steps can be carried out for different meta-learning strategies as described in Sec.~\ref{sec:meta}, with set $\mathcal{D}^{t-1}$ in lieu of the meta-training set $\mathcal{D}$ and set $\mathcal{D}_t$ for the meta-test set $\mathcal{D}_\text{T}$.
As a special case, if MAML is used, this recovers the Follow The Meta Leader (FTML) algorithm \cite{finn2019online}, which determines the shared parameter $\theta_t$ by solving the problem
\begin{align} \label{eq:ftml}
\theta_{t} = \argmin_\theta \sum_{k=1}^{t-1} L_{\mathcal{D}^\text{te}_k}(\phi_t),
\end{align}
where the context variable $\phi_t$ is computed from the local updates \eqref{eq:maml-sgd-single}--\eqref{eq:maml-sgd-multi} starting from the initial value $\theta$. The general algorithm for online meta-learning is summarized in Algorithm~\ref{alg:online-meta-learning}.

\begin{algorithm}[h] 
\DontPrintSemicolon
\smallskip
\KwIn{Data sets $\{ \mathcal{D}_t, \mathcal{D}_t^\text{data} \}$ for $t=1,2,\ldots$; step size hyperparameters $\eta$ and $\kappa$; number of transmitted pilots $p$}
\KwOut{Learned parameter vector $\theta_t$ and context vector $\phi_t$, for $t=1,2,\ldots$}
\vspace{0.15cm}
\hrule
\vspace{0.15cm}
{\bf initialize} parameter vector $\theta_1$ \\
{\bf initialize} the meta-training dataset $\mathcal{D}$ as empty, i.e., $\mathcal{D} \leftarrow [ \hspace{0.2cm} ]$ \\
\For{$t = 1,\ldots$}{
receive $\mathcal{D}_t = \{ (s_t^{(n)}, y_t^{(n)}): n=1,\ldots,p \}$\\ 
\texttt{adaptation on the current device} \\
setting $\mathcal{D}_\text{T}\leftarrow \mathcal{D}_t$ \\
follow \emph{adaptation on the meta-test device} in Algorithm~\ref{alg:maml_fomaml_reptile} or Algorithm~\ref{alg:cavia} to obtain context vector $\phi_\text{T}\rightarrow \phi_t$ \\
use $\phi_t,\theta_t$ to demodulate $\mathcal{D}^\text{data}_t$

\texttt{meta-learning phase}  \\
add current dataset $\mathcal{D}_t$ to 
meta-training dataset $\mathcal{D}$ as
$\mathcal{D} \leftarrow \bigcup^t_{k=1} \mathcal{D}_k$ \\
follow \emph{meta-learning phase} in Algorithm~\ref{alg:maml_fomaml_reptile} or Algorithm~\ref{alg:cavia} to obtain shared parameter $\theta_{t+1}$ \\
}
\caption{Few-Pilot Demodulator Online Meta-Learning}
\label{alg:online-meta-learning}
\end{algorithm}

\subsection{Integrated Online Meta-Learning and Pilot \textcolor{black}{Number Selection}}
\label{subsection:int-online-meta-and-pilot-alloc}
In order to further reduce the pilot overhead, we now consider the possibility to adapt the number of transmitted pilot symbols in each slot $t$ based on the performance of the demodulator meta-learned in the previous slots. We note that in \cite{simeone2004adaptive} the idea of adapting the number of pilots was proposed for a single device by leveraging the temporal correlation of the channels for an individual device. In contrast, the method proposed here works by using information from different devices without making any assumption about temporal correlations. \note{In practice, the adaptive assignment of the number of pilots requires a low-rate control downlink channel to be available for transmission prior to uplink transmission. The proposed method is hence not applicable to IoT systems whose communication protocols do not allow for the downlink control channels. It is noted that IoT systems such as NB-IoT do include downlink control channels \cite{wang2017primer}.}

\begin{algorithm}[h] 
\DontPrintSemicolon
\smallskip
\KwIn{Data sets $\{ \mathcal{D}_t, \mathcal{D}_t^\text{data} \}$ for $t=1,2,\ldots$; step size hyperparameters $\eta$ and $\kappa$}
\KwOut{Learned parameter vector $\theta_t$ and context vector $\phi_t$, for $t=1,2,\ldots$}
\vspace{0.15cm}
\hrule
\vspace{0.15cm}
{\bf initialize} parameter vector $\theta_1$ \\
{\bf initialize} the meta-training dataset $\mathcal{D}$ as empty, i.e., $\mathcal{D} \leftarrow [ \hspace{0.2cm} ]$ \\
{\bf initialize} number of transmitted pilots $P_1 \leftarrow P$  \\
\For{$t = 1,\ldots$}{
receive $\mathcal{D}_t = \{ (s_t^{(n)}, y_t^{(n)}): n=1,\ldots,P_t \}$ \\ 
\texttt{adaptation on the current device} \\
\For{$p = 1,\ldots,P_t$}{
setting $\mathcal{D}_\text{T} \leftarrow \{ (s_t^{(n)}, y_t^{(n)}): n=1,\ldots,p \}$ \\
follow \emph{adaptation on the meta-test device} in Algorithm~\ref{alg:maml_fomaml_reptile} or Algorithm~\ref{alg:cavia} to obtain context vector $\phi_\text{T} \rightarrow \phi_t$ \\
\uIf{(reliability check passed)}{
    set $P_{t+1}=p$ and exit \;
  }
  \uElseIf{(reliability check not passed) and ($p=P_t$)}{set $P_{t+1}=P $\;
  }
}
use $\phi_t,\theta_t$ to demodulate $\mathcal{D}^\text{data}_t$ \\
\texttt{meta-learning phase}  \\
add current dataset $\mathcal{D}_t$ to meta-training dataset $\mathcal{D}$ as
$\mathcal{D} \leftarrow \bigcup^t_{k=1} \mathcal{D}_k$ \\
follow \emph{meta-learning phase} in Algorithm~\ref{alg:maml_fomaml_reptile} or Algorithm~\ref{alg:cavia} to obtain shared parameter $\theta_{t+1}$ \\ 
}
\caption{Few-Pilot Demodulator Learning via Online Meta-Learning with Adaptive Pilot Number Selection}
\label{alg:online}
\end{algorithm}

In the proposed scheme, at each slot $t$, a device transmits $P_t$ pilots. The BS carries out demodulation of the data payload by using the demodulator $p(s|y,\phi^{(p)}_t,\theta_t)$, where the shared parameter $\theta_t$ is obtained as discussed in Sec.~\ref{subsection:online-meta} and the context variable $\phi^{(p)}_t$ is obtained by using $p\leq P_t$ pilots via Algorithm~\ref{alg:online-meta-learning}. By trying different values of $p=1,\ldots,P_t$, the BS determines the minimum value of $p\leq P_t$ such that demodulation of the data in set $\mathcal{D}^\text{data}_t$ meets some reliability requirement. If such a value of $p$ is found, then the BS assigns the number of pilots for the next slot as $P_{t+1}=p$. \textcolor{black}{Otherwise, $P_{t+1}$ is set to the maximum value $P$.} The overall online meta-learning procedure with \textcolor{black}{this pilot number selection} scheme is summarized in Algorithm~\ref{alg:online}, and an illustration of the proposed adaptive pilot number selection strategy can be found in Fig. \ref{fig:online-diag}. 

In practice, the reliability level can be estimated in different ways. For example, it can be obtained by evaluating the output of a cyclic redundancy check (CRC) field at the output of a decoder operating on the demodulated symbols from the payload $\mathcal{D}^\text{data}_t$. \textcolor{black}{Here, a simpler approach is considered that uses directly the output of the demodulator $p(s|y,\phi_t,\theta_t)$ without having to run a decoder.} This is done by comparing the cross-entropy loss \eqref{eq:loss-CE} on the demodulated data
\begin{align} \label{eq:rel-CE}
-\sum_{y\in \mathcal{D}^{\text{data}}_t} \max_{s}[{\log p(s|y,\phi^{(p)}_t,\theta_t)}]
\end{align}
to some prescribed threshold: if \eqref{eq:rel-CE} is below a threshold, then the reliability check is considered successful.

\section{Experiments}
\label{sec:exp}

In this section, we provide numerical results in order to bring insights into the advantages of meta-learning \footnote{\textcolor{black}{Code is} available at \url{https://github.com/kclip/meta-demodulator}.}.
\subsection{Offline Meta-Learning: Binary Fading}
\label{sec:exp-toy}
We begin by considering the offline set-up and focusing on a simple example, in which \note{the transmitter is ideal}, i.e., $x_k = s_k$ and fading is binary, i.e., the channel $h_k$ in \eqref{eq:e2e-channel} can take values $\pm 1$. \note{We emphasize that this set-up is intended to yield useful intuitions on the operation of meta-learning in the simplest possible setting. A more realistic scenario is studied in the next subsection.} 
\textcolor{black}{In this set-up, pulse-amplitude modulation with four amplitude levels ($4$-PAM)\note{, i.e., } $\mathcal{S} = \{-3,-1,1,3\}$, is adopted.} 
Pilot symbols in the meta-training dataset $\mathcal{D}$ and meta-test dataset $\mathcal{D_\text{T}}$ follow a fixed periodic sequence $-3,-1,1,3,-3,-1,\ldots$, while transmitted symbols in the test set for the meta-test device are randomly selected from the set $\mathcal{S}$. The channel of the meta-test device is selected randomly between $+1$ and $-1$ with equal probability, while the channels for half of the meta-training devices are set as $+1$ and for the remaining half as $-1$.

Other numerical details are as follows. The number of meta-training devices is $K = 20$; the number of pilot symbols per \note{meta-training} device is \note{$N = 1000$}, \textcolor{black}{which are divided into $N^\text{tr}=1$ training symbols and $N^\text{te}=999$ testing symbols. Only one pilot symbol is transmitted by the meta-test device, i.e. $P=1$.} The demodulator \eqref{eq:softmax-demod} is a neural network with $L=3$ layers, i.e., an input layer with $2$ neurons, one hidden layer with $30$ neurons, and a softmax output layer with $4$ neurons. The network adopts a hyperbolic tangent function $\sigma(\cdot) = \tanh(\cdot)$ as the activation function. 
\textcolor{black}{For meta-learning, fixed learning rates $\eta = 0.1$ and $\kappa = 0.001$ are used with a single local update, i.e., $m = 1$, and the ADAM optimizer \cite{kingma2014adam} is used to update the shared parameter $\theta$. To train for the meta-test device, one adaptation step is adopted, i.e., $I_\text{T}=1$, with learning rate $\eta = 0.1$. The \textcolor{black}{average} signal-to-noise ratio (SNR) per real symbol is given as $2E_x/N_0 = 18\text{ dB}$.}

We compare the performance of the proposed meta-learning approach via MAML, FOMAML, REPTILE, and CAVIA with: (\emph{i}) \note{conventional learning} where data from the meta-training devices \note{are} not used \note{and the weights of the demodulator are randomly initialized}; (\emph{ii}) joint training with the meta-training dataset $\mathcal{D}$ as explained in Sec.~\ref{subsection:joint-training}; (\emph{iii}) optimal ideal demodulator that assumes perfect \note{CSI}. \note{In order to improve the performance of (\emph{i}), instead of single adaptation step for the meta-test device, we allow for multiple SGD updates with learning rate $0.001$; while, for} (\emph{ii}), we set the learning rate to $0.001$ and the mini-batch size to $4$. The probability of \note{symbol} error of \note{the optimal demodulator} (\emph{iii}) can be computed as $P_e = 3/2Q(\sqrt{\text{SNR}/5})$ using standard arguments.

\note{
\begin{figure}
    \centering
    \hspace*{-0.02in}
    \includegraphics[width=0.95\columnwidth]{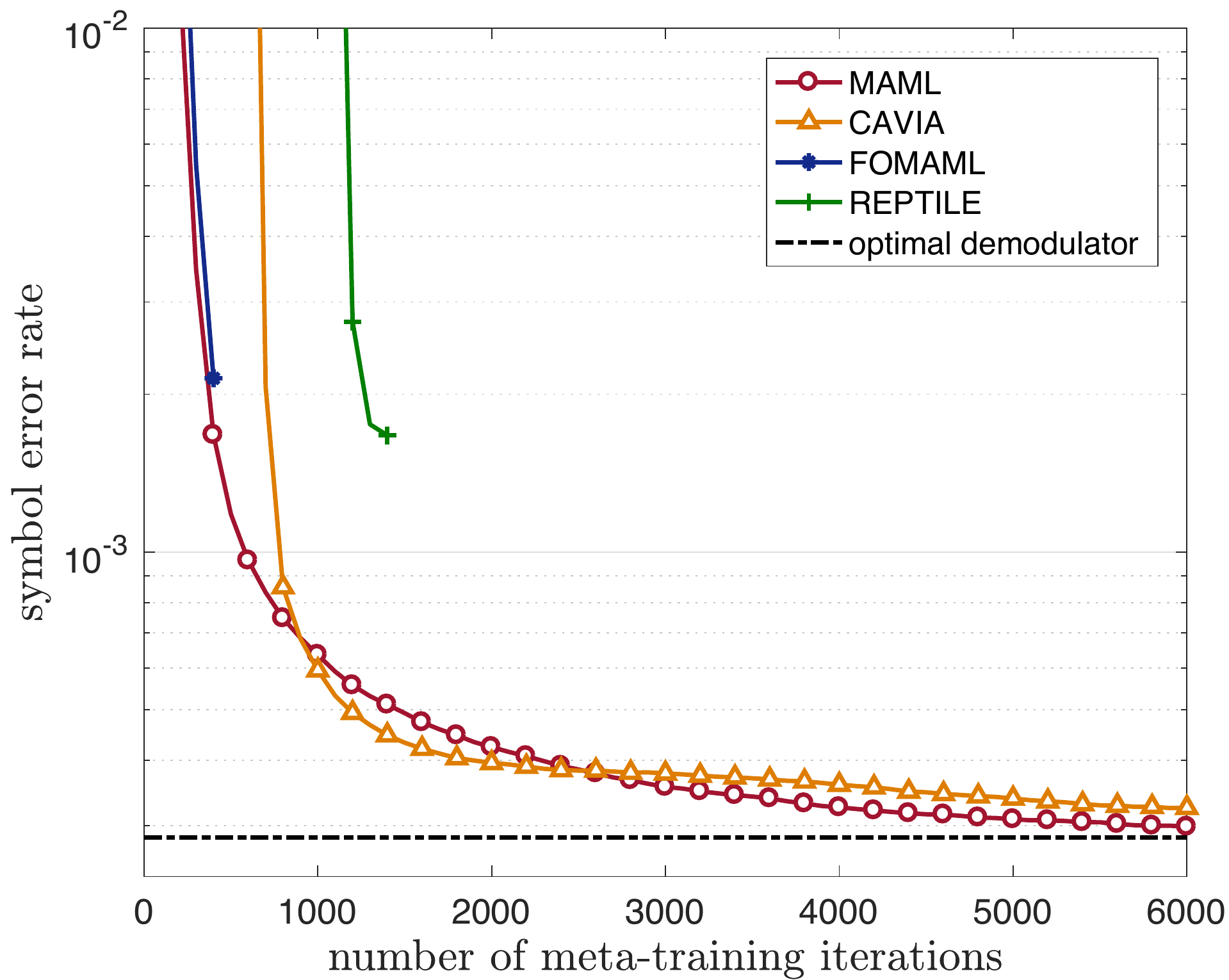}
    \caption{\note{Symbol error rate with respect to number of iterations during meta-training for an offline meta-learning example with binary fading. $P=1$ pilot is used for meta-test device. The symbol error rate is averaged over $10^6$ data symbols and $100$ meta-test devices. The unstable behavior of FOMAML and REPTILE after $400$ and $1400$ iterations, respectively, are not shown. \textcolor{black}{The symbol error rate for conventional training and joint training are not shown as their symbol error rate is above $0.25$.}}}
    \label{fig:prob_of_error_simple}
\end{figure}
}

\note{
\begin{figure}[t]
    \centering
    \hspace{-0.5cm}
    \includegraphics[width=0.8
    \columnwidth]{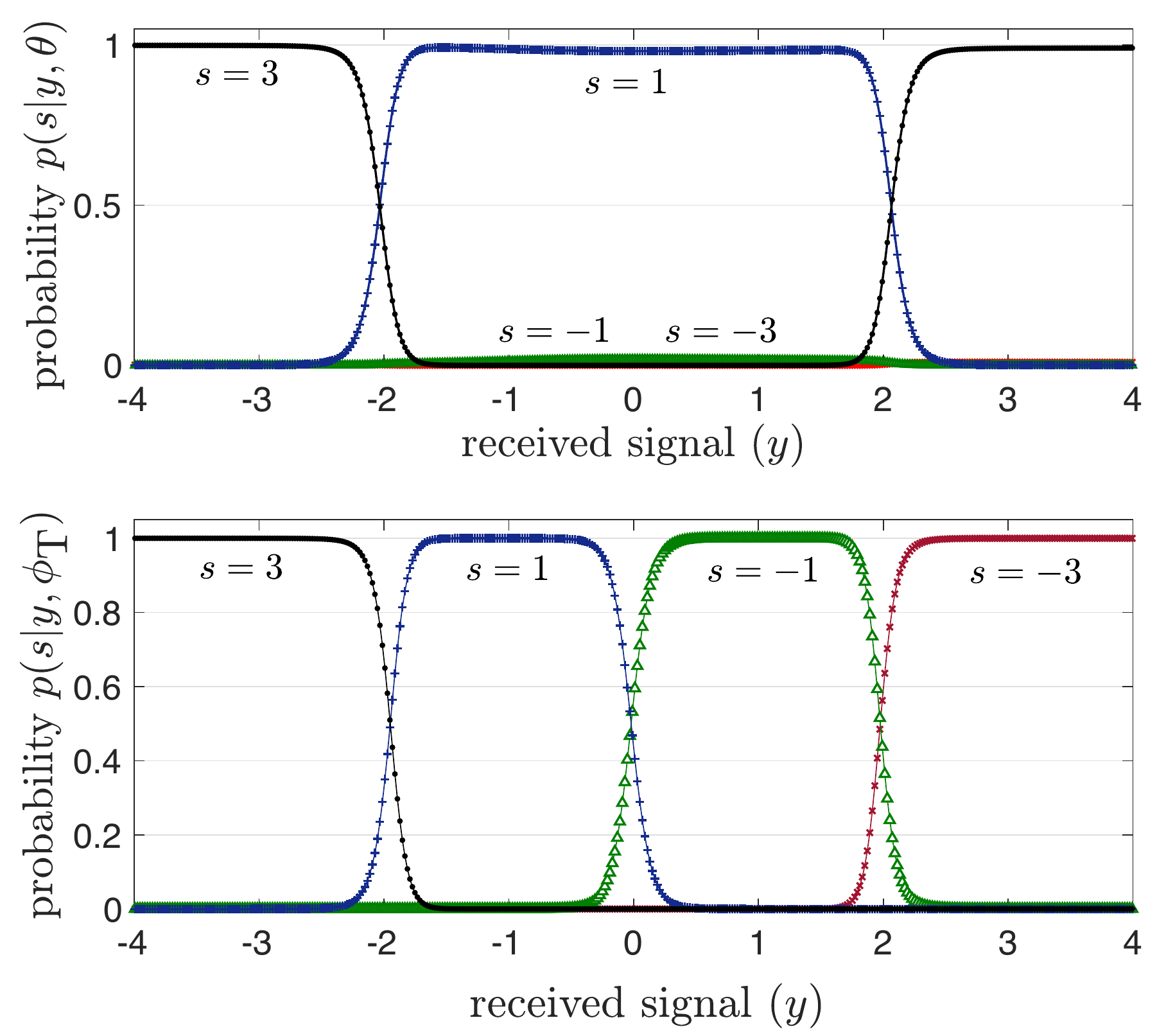}
    \caption{(Top) Demodulator \eqref{eq:softmax-demod} for the shared parameter vector $\theta$ obtained via offline meta-learning phase in Algorithm~\ref{alg:maml_fomaml_reptile} using MAML; (Bottom) \textcolor{black}{Updated demodulator \eqref{eq:softmax-demod}} with target-device specific parameter vector \textcolor{black}{$\phi_\text{T}$} using \note{$P = 1$ pilot} from the meta-test device.}
    \label{fig:vis_simple}
\end{figure}
}

\note{In Fig.~\ref{fig:prob_of_error_simple}, we plot the symbol error rate with respect to number $I$ of iterations during meta-training. Conventional learning and joint training curves are not shown since they both failed to reach symbol error rates smaller than $0.25$. FOMAML and REPTILE curves are shown up to $400$ and $1400$ meta-training iterations, respectively, since further iterations for both schemes show unstable behavior \cite{antoniou2018train, keskar2017improving}. Despite having only one pilot for adaptation $P=1$, both MAML and CAVIA can approach the performance of the optimal demodulator, with MAML outperforming CAVIA after sufficiently many meta-training iterations. This adavantage of MAML stems from its adaptation of a larger number of parameters $\phi_\text{T}$, namely the demodulator weight vector, with respect to CAVIA, which only updates an auxiliary input vector.} Overall, these results confirm the claim that, unlike conventional solutions, meta-training can effectively transfer information from meta-training devices to a new target device \note{to achieve near-optimal demodulator with few pilots, here only one.}

In order to gain intuition on how meta-learning learns from the meta-training devices, in Fig.~\ref{fig:vis_simple}, we plot the probabilities defined by the demodulator \eqref{eq:softmax-demod} for the four symbols in the constellation $\mathcal{S}$ with the shared parameter vector $\theta$ obtained from the meta-learning phase in Algorithm~\ref{alg:maml_fomaml_reptile} (top) and with target-device specific parameter vector $\phi_{\text{T}}$ after adaptation using the pilots of the target meta-test device (bottom). \textcolor{black}{Here, MAML is adopted as the meta-learning algorithm.} The class probabilities identified by meta-learning in the top figure have the interesting property of being approximately symmetric with respect to the origin. \note{From this symmetric initialization,} the resulting decision region \note{can be adapted} to the channel of the target device, which may take values $\pm 1$ in this example. The adapted probabilities in the bottom figure illustrate how the initial demodulator obtained via MAML is specialized to the channel of the target device.

\subsection{Offline Meta-Learning: Rayleigh Fading \note{with I/Q Imbalance}}
\label{sec:exp-real}

\begin{figure}
    \centering
    \hspace{-0.5cm}
    \includegraphics[width=0.93\columnwidth]{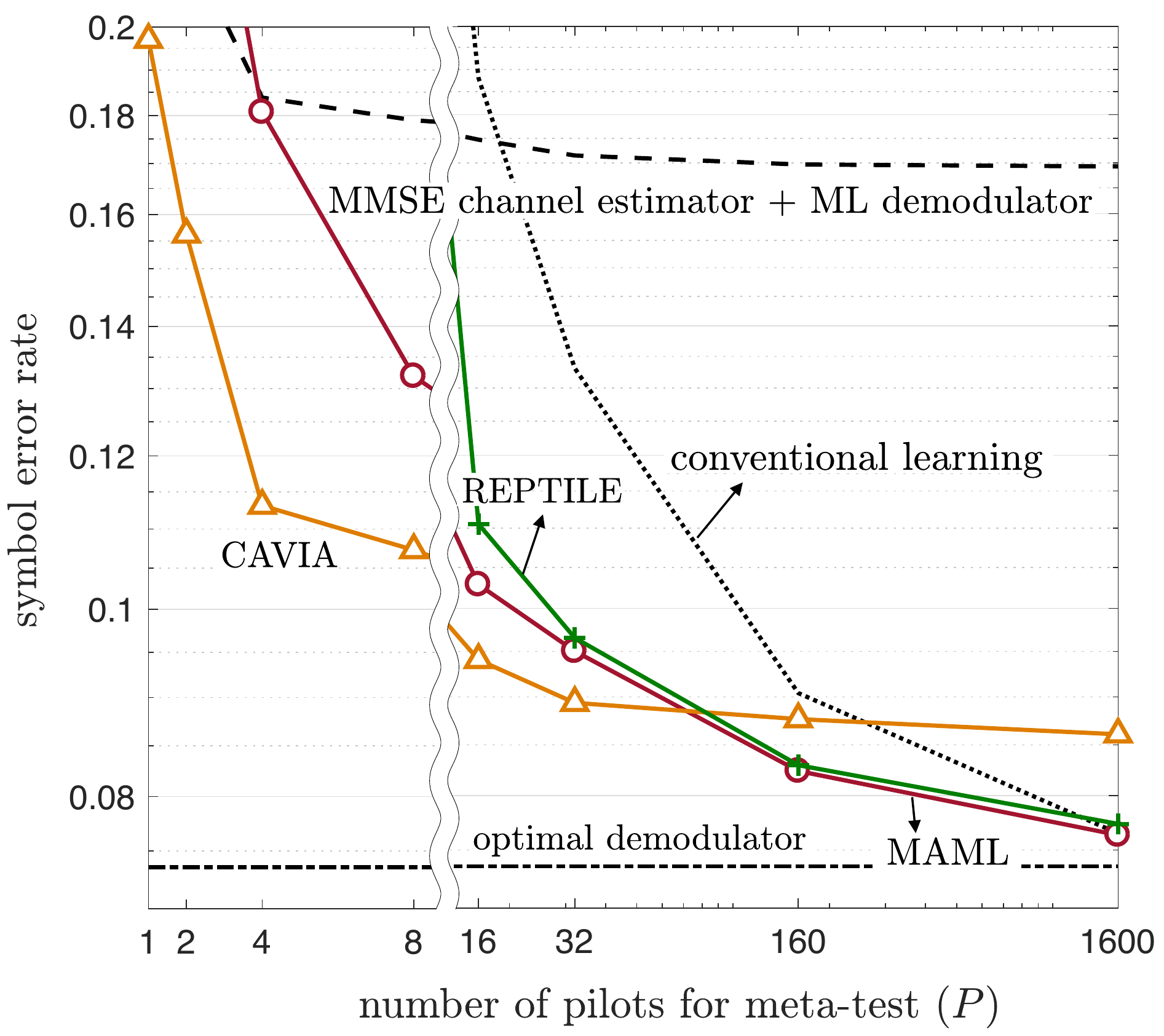}
    \caption{\note{Symbol error rate with respect to \textcolor{black}{the} number $N^\text{tr} = P$ of training pilots used for both meta-training and meta-testing for offline meta-learning with $16$-QAM, Rayleigh fading, and I/Q imbalance with $K=1000$ meta-training devices, $N^\text{tr}+N^\text{te}=3200$ pilots for meta-training devices. In order to plot average \textcolor{black}{extra} symbol error probability on the meta-test device, the symbol error rate is evaluated by averaging over $100$ different meta-test devices (i.e., channel realizations), each with $10000$ data symbols.}}
    \label{fig:off_real_1}
\end{figure}
We now consider a more realistic scenario including Rayleigh fading channels $h_k\sim\mathcal{CN}(0,1)$ and model  \eqref{eq:iq_imbalance} to account for \note{I/Q imbalance at the transmitters. We set $\epsilon_k=0.15 \epsilon'_k$ and $\delta_k=15^\circ \delta'_k$, where $\epsilon'_k$ and $\delta'_k$ are independent random variables with beta distribution $\text{Beta}(5,2)$. Note that this implies that $\epsilon_k$ and $\delta_k$ are limited in the intervals $[0,0.15]$ and $[0,15^\circ]$, respectively.} We assume $16$-ary quadrature amplitude modulation ($16$-QAM) for constellation $\mathcal{S}$\note{,} and the sequence of pilot symbols in the meta-training dataset $\mathcal{D}$ and meta-test dataset $\mathcal{D_\text{T}}$ \note{is} fixed by cycling through the symbols in $\mathcal{S}$, while the transmitted symbols in the test set for the meta-test device are randomly selected from $\mathcal{S}$. 
The number of meta-training devices is set as \note{$K=1000$}; the number of pilot symbols per device is \note{$N=3200$, which are divided into $N^{\text{tr}}$ training samples and $N^{\text{te}}=N-N^\text{tr}$ testing samples. The \textcolor{black}{average} SNR per complex symbol is given as $E_x/N_0 = 20\text{ dB}$.} Further details on the numerical set-up can be found in Appendix~\ref{sec:appdendix-set-up}.

In Fig.~\ref{fig:off_real_1}, \textcolor{black}{the symbol error rate with respect to the number $P$ of pilots for the meta-test device is illustrated when using an equal number of pilots for meta-training, i.e., $N^\text{tr}=P$.} As in Fig.~\ref{fig:prob_of_error_simple}, we compare the performance of meta-training methods with \note{conventional learning} and joint training strategies, \note{along with a conventional communication scheme based on Minimum Mean Square Error (MMSE) channel estimation with $P$ pilots followed by a Maximum Likelihood (ML) demodulator. All the schemes with worse performance as compared to this conventional communication scheme are not shown. MAML, REPTILE, and CAVIA are seen to adapt quickly to the channel and I/Q imbalance of the target device, outperforming the conventional scheme based on MMSE channel estimation, which is agnostic to the I/Q imbalance.} MAML shows the best performance for sufficiently large $P$, while CAVIA \note{is seen to outperform MAML when} fewer pilots \note{are available}. \note{It is worth noting that, when there is a sufficient number $P$ of pilots for the meta-test device, conventional learning can outperform meta-learning schemes. In this case, the inductive bias inferred by meta-training can hence cause a performance degradation \cite{park2019meta}.}

\begin{figure}
    \centering
    \hspace{-0.5cm}
    \includegraphics[width=0.91\columnwidth]{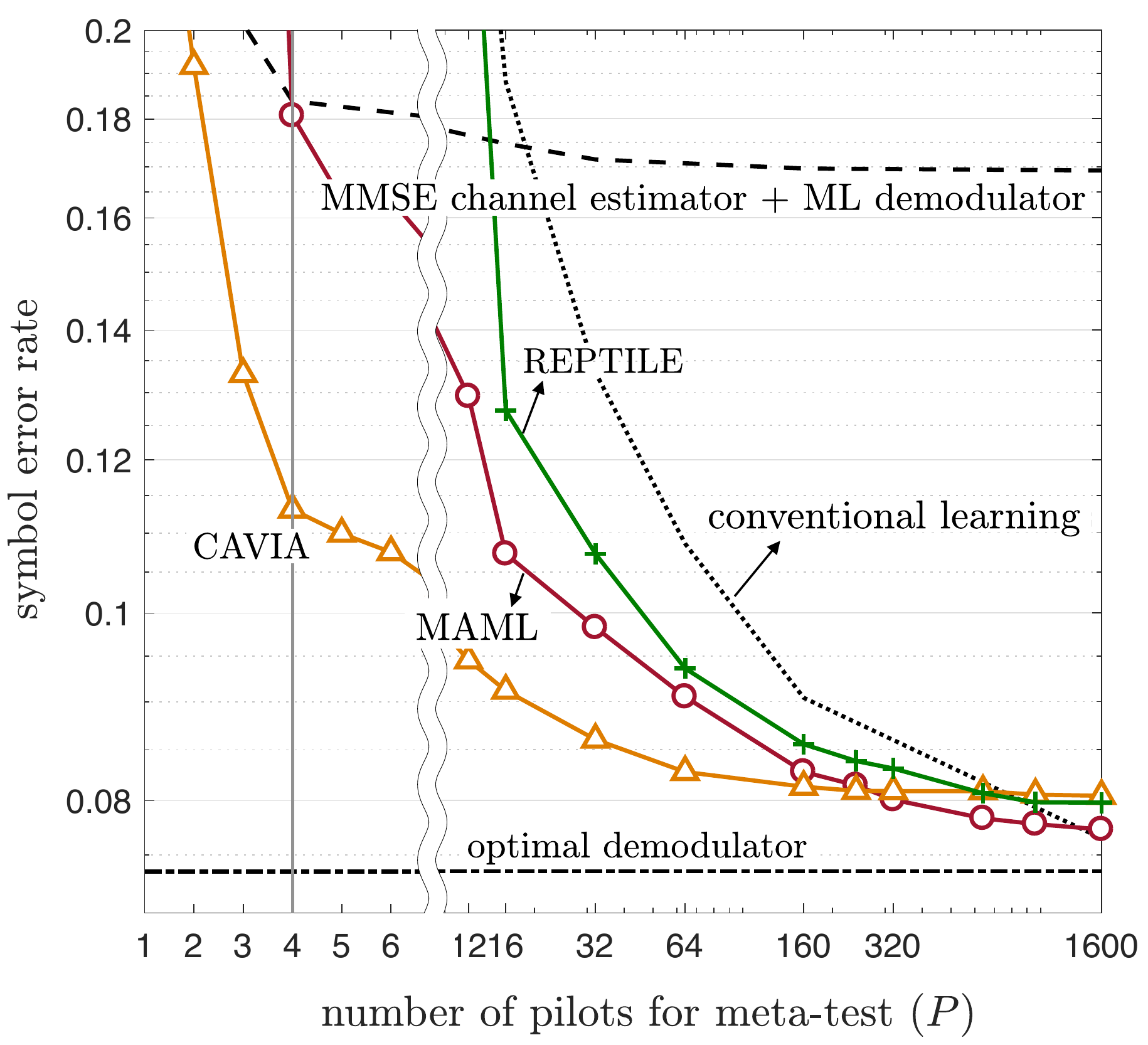}
    \caption{\note{Symbol error rate with respect to \textcolor{black}{the} number $P$ of pilots (used during meta-testing) for offline meta-learning with $16$-QAM, Rayleigh fading, and I/Q imbalance for $K=1000$ meta-training devices, $N^\text{tr}=4$ (vertical line), $N^\text{te}=3196$. The symbol error is averaged \textcolor{black}{over $10000$} data symbols and $100$ meta-test devices.}}
    \label{fig:off_real_2}
\end{figure}

\note{In Fig.~\ref{fig:off_real_2}, we consider the case where \textcolor{black}{the} number $P$ of pilots for the meta-test device is different from the number $N^\text{tr}$ used for meta-training, here set to $N^\text{tr}=4$. Despite this mismatch between meta-training and meta-testing condition, MAML, CAVIA, and REPTILE are seen to outperform conventional communication when there is a sufficient number $P$ of pilots for meta-test. In a manner similar to results in Fig. ~\ref{fig:off_real_1}, CAVIA shows the best performance with extremely few pilots, \textcolor{black}{e.g., $4$ pilots}, while MAML is preferable for larger values of $P$. \note{In fact, comparing Fig.~\ref{fig:off_real_1} and Fig.~\ref{fig:off_real_2} reveals that having $P>N^\text{tr}$ can even be advantageous for some meta-training schemes, such as CAVIA. This may be interpreted in terms of meta-overfitting, which refers to a degradation in meta-testing performance due to an excessive dependence of the meta-trained shared parameters on the meta-training data \cite{amit2017meta}. Using fewer pilots during meta-training can potentially reduce meta-overfitting, by making the shared parameters less dependent on meta-training data, and improve meta-testing performance.}}

\begin{figure}
    \centering
    \hspace{-0.5cm}
    \includegraphics[width=0.9\columnwidth]{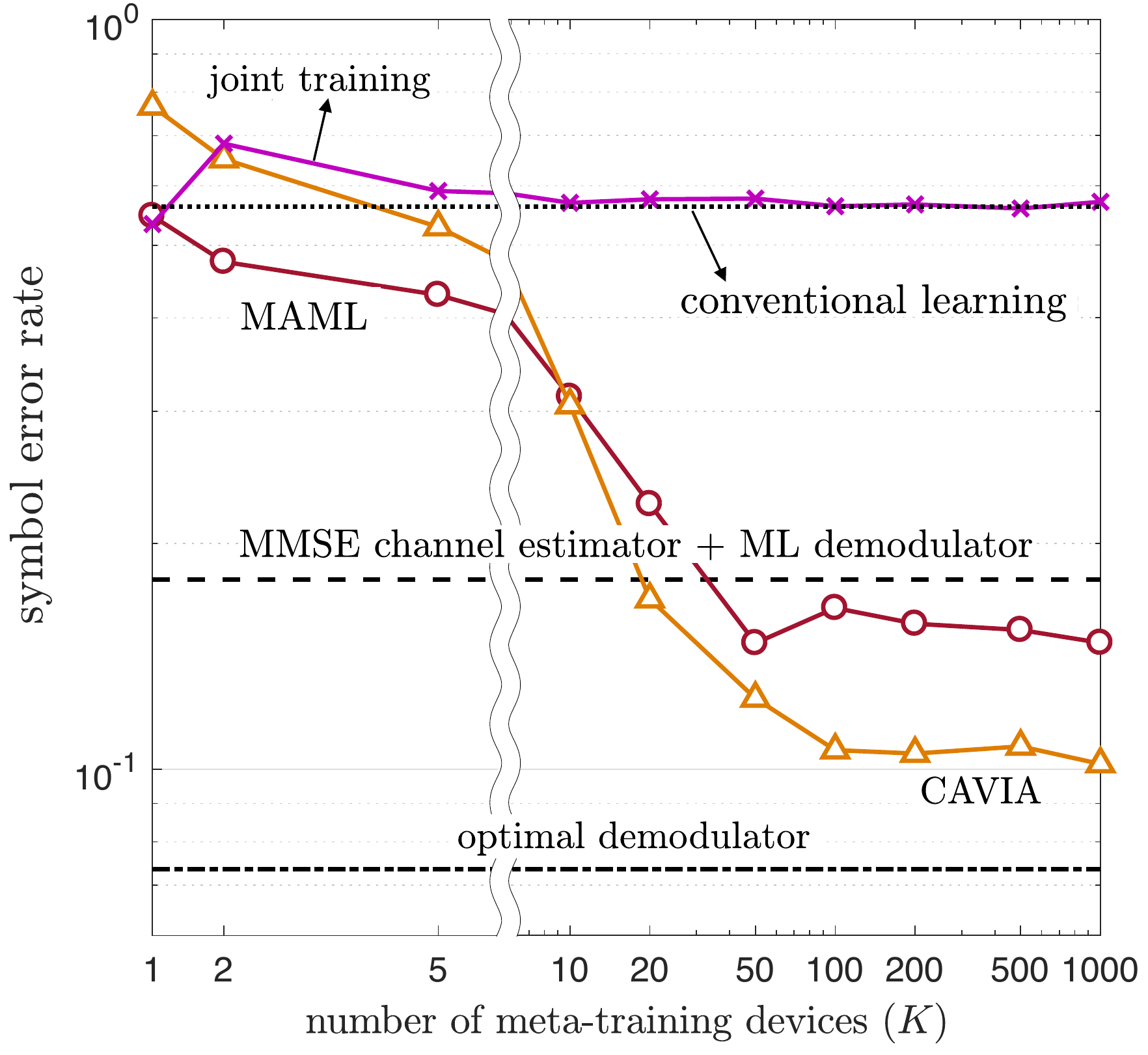}
    \caption{\note{Symbol error rate with respect to number $K$ of meta-training devices for offline meta-learning with $16$-QAM, Rayleigh fading, and I/Q imbalance with $N^\text{tr}=4$ and $N^\text{te}=3196$ for meta-training devices, $P=8$ pilots for meta-test devices. The symbol error is averaged over by $10000$ data symbols and $100$ meta-test devices.}}
    \label{fig:meta-device}
\end{figure}

Finally, in Fig.~\ref{fig:meta-device}, \textcolor{black}{the symbol error rate with respect to the number $K$ of the meta-training devices is demonstrated}. \note{Joint training has a performance similar to conventional learning, hence being unable to transfer useful information from the $K$ meta-training devices. In contrast, MAML and CAVIA} show better performance when given \note{data from} more meta-training devices, up to a point where the gain saturates. This matches well with the intuition that there is only a limited amount of common information among different users that can be captured by meta-learning. Confirming the results in Fig.~\ref{fig:off_real_1} and Fig.~\ref{fig:off_real_2}, MAML and CAVIA are seen to offer \note{better} performance \note{than the conventional communication scheme with} a sufficient number $K$ of meta-training devices. Furthermore, CAVIA needs a larger value of $K$ than MAML. This accounts again for CAVIA's architectural difference as compared to MAML: CAVIA needs to find a shared parameter vector $\theta$ for the demodulator $p(s|y,\phi_\text{T},\theta)$ that is not adapted to the training symbols of the current device.


\subsection{Online Meta-Learning\note{: Rayleigh Fading with I/Q Imbalance}}
\label{sec:exp-online}
We now move on to consider the online scenario under same assumptions on Rayleigh fading, transmitters' I/Q imbalance, modulation scheme, and SNR as in the offline set-up presented in Sec.~\ref{sec:exp-real}. 

The maximum number of pilots is set as $P=32$, and 
\textcolor{black}{the number of pilots $P_t$ in any slot $t$ is determined by using adaptive pilot number selection scheme in Algorithm~\ref{alg:online}.} 
In a manner similar to the offline set-up, we compare the performance with: \emph{(i)} a \note{conventional learning} scheme that only adapts to current device based on current pilot data $\mathcal{D}_t$ with number of pilots $P_t$ fixed as constant to a prescribed value; \emph{(ii)} joint training as described in Sec.~\ref{subsection:online-learning}\note{; \emph{(iii)} conventional communication scheme with MMSE channel estimation and ML demodulator; and \emph{(iv)} optimal ideal demodulator as described in the previous section.} Other details on the numerical set-up can be found in Appendix~\ref{sec:appdendix-set-up}.

\begin{figure}
    \centering
    \hspace{-0.65cm}
    \includegraphics[width=0.9\columnwidth]{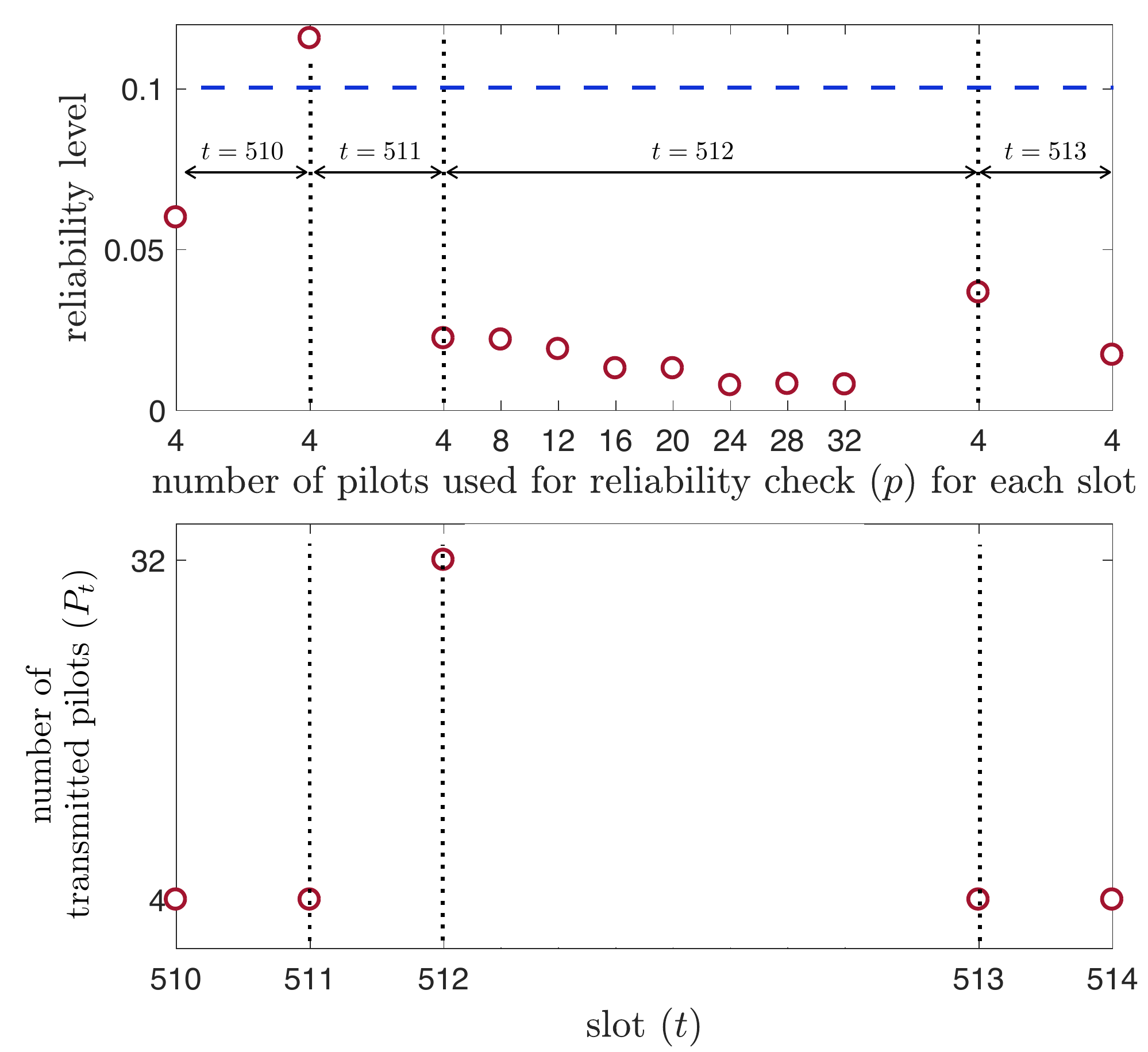}
    \caption{Illustration of the procedure of adaptive pilot number selection scheme: (top) reliability level in \eqref{eq:rel-CE} versus the number $p$ of pilots during slots \note{$t=510,\ldots,513$} (the prescribed reliability threshold value, \note{$0.1$}, is dashed); (bottom) number of transmitted pilots $P_t$ for each slot \note{$t=510,\ldots,514$}.}
    \label{fig:online-1}
\end{figure}

In Fig.~\ref{fig:online-1}, we first describe the procedure used by the proposed adaptive pilot number selection scheme. As discussed in Sec.~\ref{subsection:int-online-meta-and-pilot-alloc}, \textcolor{black}{reliability levels for different values of the number $p$ of pilots are evaluated by using \eqref{eq:rel-CE} as shown in Fig.~\ref{fig:online-1} (top)} 
and the number of transmitted pilots $P_{t+1}$ in the next slot is selected accordingly (bottom). The adaptive pilot number selection scheme is performed here with \note{CAVIA}, while the prescribed threshold value is set as \note{$0.1$}. For instance, for slot \note{$513$}, the number of transmitted pilots is chosen as \note{$P_{513}=4$} based on the result from previous slot \note{$512$} that passed reliability check at \note{$p=4$}. In contrast, for slot \note{$512$}, the number of transmitted pilots \note{$P_{512}=P$} has been chosen as maximum value $p=32$ due to the failure of reliability check pass at slot \note{$511$}. In the following, we assess whether the adaptive pilot number selection scheme can maintain reasonable performance in terms of probability of symbol error in the payload data $\mathcal{D}_t^\text{data}$, despite the illustrated reduction in the pilot overhead.

\begin{figure}
    \centering
    \hspace{-0.5cm}\includegraphics[width=0.92\columnwidth]{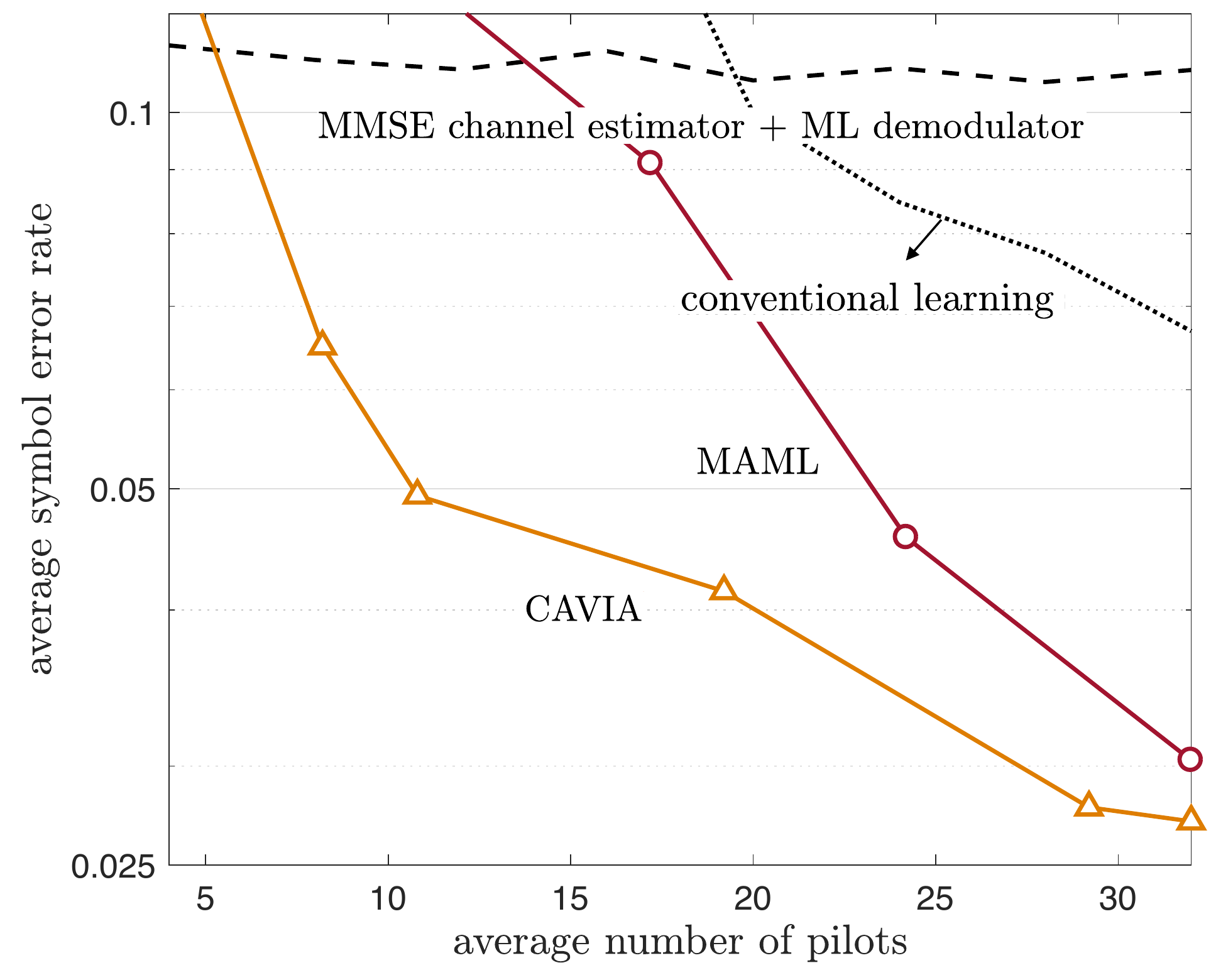}
    \caption{Average symbol error rate with respect to average number of pilots over slots \note{ $t=500,\ldots,520$} for online meta-learning.}
    \label{fig:online-2}
\end{figure}

To this end, in Fig.~\ref{fig:online-2}, we plot the average symbol error rate for payload data $\mathcal{D}_t^\text{data}$ versus the average number of transmitted pilots as evaluated \note{during slots} \note{ $t=500,\ldots,519$.} For \note{MAML and CAVIA}, the corresponding curve is obtained by selecting the threshold values \note{$(0, 0.01, 0.05, 0.1)$ and $(0, 0.001, 0.02, 0.05, 0.1, 0.15)$}, respectively, for the reliability level. For \note{conventional learning}, each point on the curve corresponds to the given fixed number of pilots defined by the horizontal axis. The proposed adaptive pilot number selection scheme is seen to improve \note{both over conventional learning and a conventional model-based demodulator}. In particular, in a manner consistent with the discussion so far, \note{MAML and CAVIA can operate with fewer pilots than conventional schemes, with CAVIA being generally more efficient in reducing the number of pilots due to its adaptation of a smaller number of parameters.}

\begin{table*}[bp]
\caption{\note{Numerical Set-Up}} \label{tab:title} 
\centering
\begin{tabular}{ |p{9cm}||p{1.8cm}|p{2.3cm}|p{2.6cm}|  }
 \hline
 & \note{Figs. 6, 7}  & \note{Figs. 8, 9, 10 } & \note{Figs. 11, 12}\\
 \hline
 \note{number of meta-training iterations ($I$)}   & \note{-}   &\note{50000}&   \note{500 (at each time slot)}\\
\note{number of local updates during meta-training ($m$)}&   \note{1}  & \note{1}   &\note{1}\\
 \note{number of local updates during meta-testing ($I_\text{T}$) }&\note{1} & \note{1000}&  \note{1000}\\
 \note{mini-batch size during meta-training}  &\note{$N^\text{tr}=1$} & \note{$N^\text{tr}$}&  \note{$N^\text{tr}=4$}\\
 \note{mini-batch size during meta-testing for $i_\text{T}$th update, for $i_\text{T}=1,\ldots,m$}&   \note{$1$}  & \note{$\min(P,N^\text{tr})$ }& \note{$\min(P,N^\text{tr})$} \\
\note{ mini-batch size during meta-testing for $i_\text{T}$th update, for $i_\text{T}=m+1,\ldots,I_\text{T}$  }&   \note{-}  & \note{$\min(P,16)$} & \note{$\min(P,16)$}\\
  \note{learning rate $\eta$ during meta-training} & \note{0.1} &\note{0.1} &\note{0.1} \\
  \note{learning rate $\kappa$ during meta-training} & \note{0.001} &\note{0.001} &\note{0.001} \\
  \note{learning rate $\eta$ during meta-testing for $i_\text{T}$th update, for $i_\text{T}=1,\ldots,m$}& \note{0.1}  & \note{0.1}   &\note{0.1}\\
 \note{learning rate $\eta$ during meta-testing for $i_\text{T}$th update, for $i_\text{T}=m+1,\ldots,I_\text{T}$} & \note{-} &  \note{0.005}&\note{0.005}\\
 \hline
\end{tabular}
\end{table*}  
\vspace{0.2cm}
\section{Conclusions and Extensions}
\label{sec:conclusions-future-works}

In communication systems with short packets, such as IoT, meta-learning techniques can adapt quickly based on few training examples by transferring knowledge from previously observed pilot information from other devices. 
In this paper, we have proposed the use of offline and online meta-learning for IoT scenarios by adapting state-of-the-art meta-learning schemes, namely MAML, FOMAML, REPTILE, and CAVIA, in a unified framework. 
\textcolor{black}{For the online setting, meta-learning has been further integrated with an adaptive pilot number selection scheme to reduce the pilot overhead.} 
Extensive numerical results have validated the advantage of meta-learning in both offline and online cases as compared to conventional machine learning schemes. Moreover, comparisons among the mentioned meta-learning schemes reveal that MAML and CAVIA are preferable, with each scheme outperforming the other in different regimes in terms of amount of available meta-training data. 
\textcolor{black}{For online meta-learning, the proposed pilot selection scheme was demonstrated to have a decreased pilot overhead with negligible performance degradation of the demodulator.} 

Meta-learning, first introduced in the conference version \cite{park2019learning} of this work and in \cite{jiang2019mind} for use in communications systems, may be useful in a number for other network functionalities characterized by reduced overhead and correlation across successive tasks. Examples include prediction of traffic from sets of IoT devices, e.g., in grant-free access \cite{kassab2019information}, \cite{jiang2019online}; channel estimation \cite{mao2019metaicc}; and precoding in multi-antenna systems. 

\note{A complementary approach for reducing the pilot overhead involves model-based compressive sensing (CS) methods that leverage sparsity of the channel in some domain, typically space or multi-path dimensions \cite{rao2015compressive, gao2015spatially}. An interesting direction for future work would be to combine meta-learning with CS in order to further reduce the number of pilots for \textcolor{black}{frequency-selective} or multi-antenna channels.} Furthermore, more advanced meta-training solutions can also be considered that are based on a probabilistic estimate of the context variables \cite{nguyen2019uncertaintybayes}.
Finally, this work may motivate the development of novel meta-training techniques that reap the complementary benefits of CAVIA and MAML.

\appendices
\section{Hessian-Vector Product Calculation}
\label{sec:appdendix-hess-vec-approx}
In order to compute the updates in \eqref{eq:maml-1} and \eqref{eq:maml-2}, we adopt a finite difference method for Hessian-vector product calculation \cite{lecun2012efficient}. This allows us to avoid computing Hessian matrix \note{and obtain} an approximate value of the product of the Hessian matrix and a vector. Given a loss function $L(\theta)$ defined and doubly continuously differentiable over a local neighborhood of the value $\theta$ of interest, the finite difference method approximates the Hessian-vector product $Hg$, where $H=\nabla^2_\theta L(\theta)$ is the Hessian matrix and $g$ is any vector. The Hessian-vector product $Hg$ can be approximately computed as \cite{lecun2012efficient}
\begin{align} \label{eq:appdx-hess-vec-approx}
Hg \approx \frac{1}{\alpha} (\nabla_{\theta}L(\theta+\alpha g)-\nabla_{\theta}L(\theta)),
\end{align}
where $\alpha$ is a sufficiently small constant value. In \eqref{eq:appdx-hess-vec-approx}, we follow \cite{andrei2009accelerated} to choose $\alpha$ as
\begin{align} \label{eq:appdx-hess-vec-alpha}
\alpha = \frac{2\sqrt{\epsilon}(1+\norm{\theta})}{\norm{g}},
\end{align}
where $\norm{\cdot}$ indicates Euclidean norm and $\epsilon = \num{1.1920929e-7}$, which is an upper bound on the relative error due to rounding in single precision floating-point arithmetic \cite{goldberg1991every}.

\section{Details on Numerical Set-Up}
\label{sec:appdendix-set-up}
\subsection{Offline Meta-Learning}


The following is the fixed sequence that is used for pilot symbols in the meta-training dataset $\mathcal{D}$ and meta-test dataset $\mathcal{D}_\text{T}$ in the offline scenario \note{for 16-QAM: $-3-3j, -3+1j, 1+1j, 1-3j, 
-3+3j, 3+1j, 1-1j, -1-3j,
3+3j, 3-1j, -1-1j, -1+3j,
3-3j, -3-1j, -1+1j, 1+3j, -3-3j, -3+1j,\ldots$.
Pilots in meta-test dataset $\mathcal{D}_\text{T}$ and training set $\mathcal{D}_k^\text{tr}$ follow this same fixed sequence.} For the experiments in Sec.~\ref{sec:exp-real},
every \note{demodulator} \eqref{eq:softmax-demod} except for CAVIA is a neural network with $L=5$ layers, i.e., an input layer with $2$ neurons, three hidden layer\note{s} with $10, 30, 30$ \note{neurons}, and a softmax output layer with $16$ neurons. For CAVIA, we use a neural network with an input layer of $12$ neurons, three hidden layer\note{s} with $10, 30, 30$ \note{neurons}, and a softmax output layer with $16$ neurons, so that the dimension of the context parameter $\phi$ is 10. For the activation function, we adopt a \note{ReLU function as} $\sigma(\cdot)=\text{ReLU}(\cdot)$. \note{Detailed settings are described in Table~\ref{tab:title}}.

\subsection{Online Meta-Learning}
We trained the demodulator \eqref{eq:softmax-demod} \note{with the same architecture}
with same mini-batch sizes and learning rates described above for offline meta-learning. \note{For this experiment, we set $\mathcal{D}_k^\text{te} = \mathcal{D}_k$. Other details are described in Table~\ref{tab:title}.}

\newpage

\vspace{0.2cm}
\bibliographystyle{IEEEtran}

\begin{IEEEbiography}[{\includegraphics[width=1in,height=1.25in,clip,keepaspectratio]{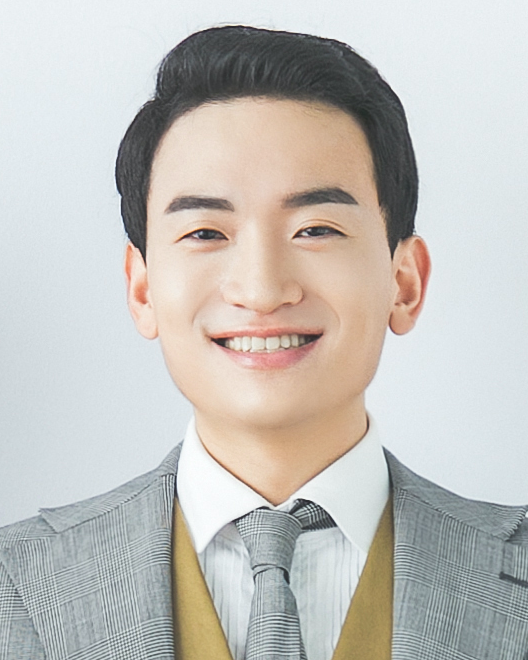}}]{Sangwoo Park} \textcolor{black}{(Student Member, IEEE) received his B.S. degree in physics in 2014 and the M.S.E degree in electrical engineering in 2016, all from Korea Advanced Institute of Science and Technology (KAIST), Daejeon, Korea. He is currently working toward the Ph.D. degree at KAIST. His research interests lie in the field of communication theory, machine learning, and statistical signal processing, with emphasis on meta-learning usage for communication systems. From April to December 2019, he was a visiting researcher at King's Communications, Learning and Information Processing lab, King's College London, United Kingdom. He has served as a Reviewer for many journals, including the IEEE Journal on Selected Areas in Communications and the IEEE Transactions on Cognitive Communications and Networking.}
\end{IEEEbiography}

\begin{IEEEbiography}[{\includegraphics[width=1in,height=1.25in,clip,keepaspectratio]{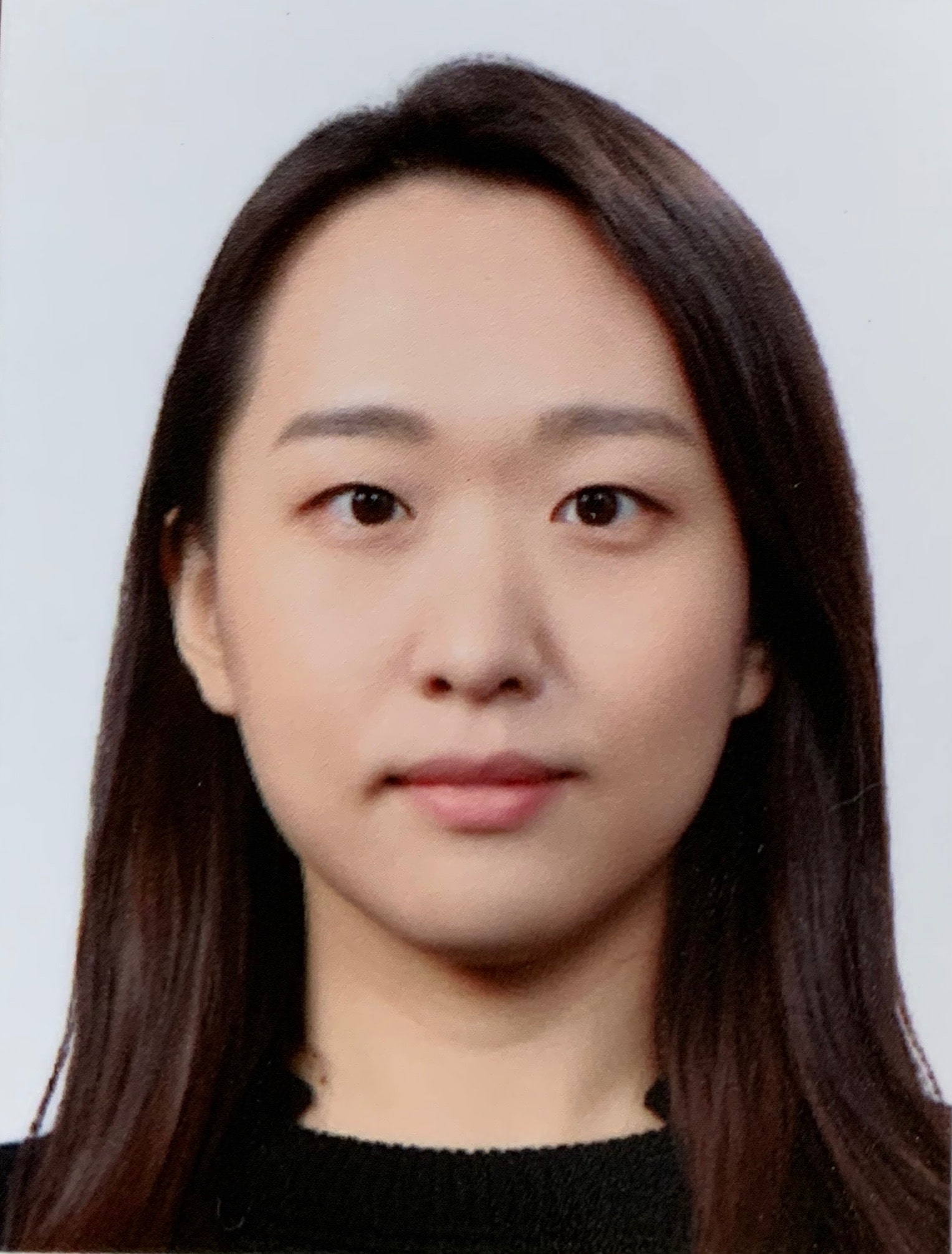}}]{Hyeryung Jang} \textcolor{black}{(Member, IEEE) received her B.S., M.S., and Ph.D. degrees in electrical engineering from the Korea Advanced Institute of Science and Technology, in 2010, 2012, and 2017, respectively. She is currently a research associate in the Department of Informatics, King’s College London, United Kingdom. Her recent research interests lie in the mathematical modeling, learning, and inference of probabilistic graphical models, with a specific focus on spiking neural networks and communication systems. Her past research works also include network economics, game theory, and distributed algorithms in communication networks.}
\end{IEEEbiography}

\begin{IEEEbiography}[{\includegraphics[width=1in,height=1.25in,clip,keepaspectratio]{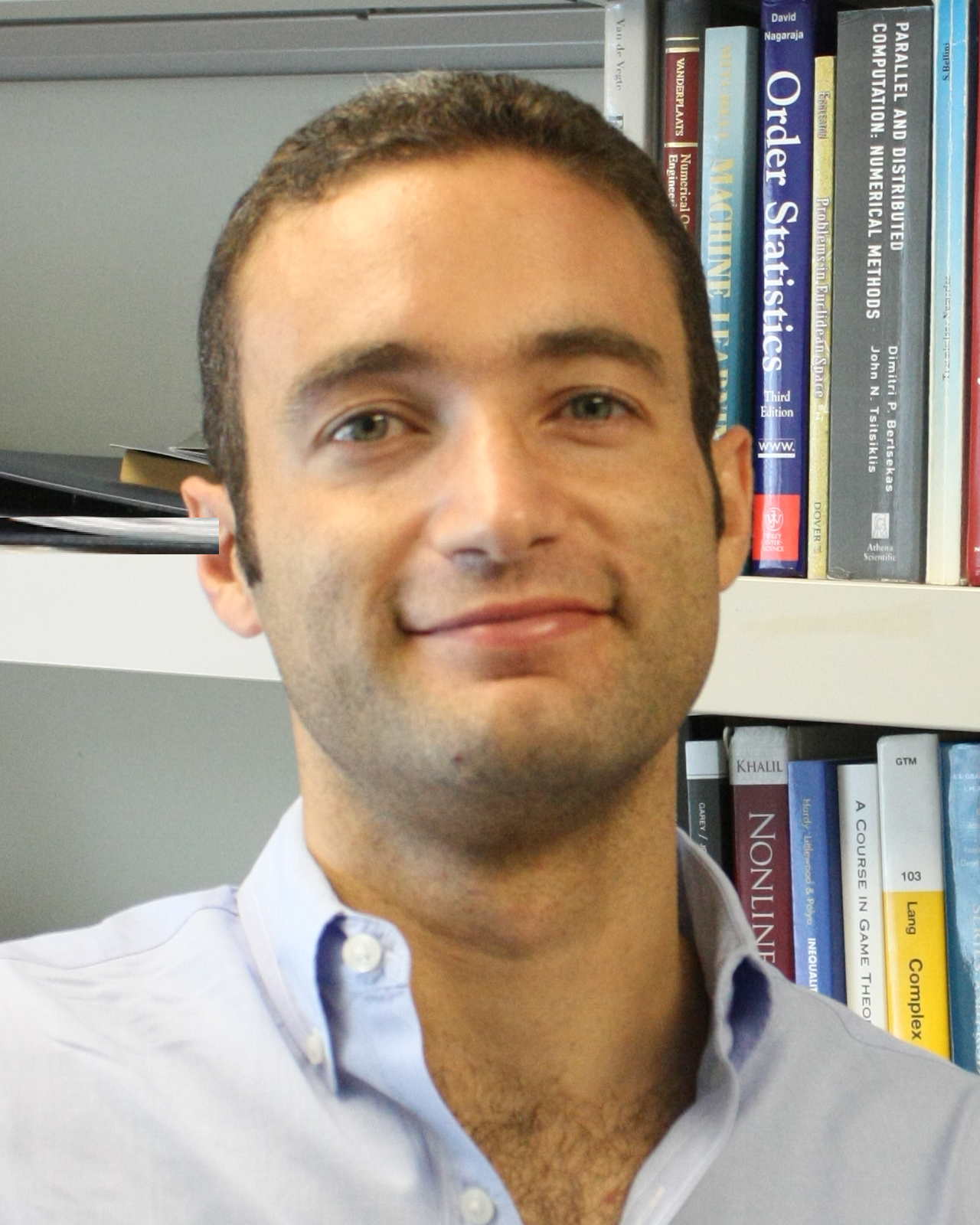}}]{Osvaldo Simeone}\textcolor{black}{(Fellow, IEEE) is a Professor of Information Engineering with the Centre for Telecommunications Research at the Department of Engineering of King's College London, where he directs the King's Communications, Learning and Information Processing lab. He received an M.Sc. degree (with honors) and a Ph.D. degree in information engineering from Politecnico di Milano, Milan, Italy, in 2001 and 2005, respectively. From 2006 to 2017, he was a faculty member of the Electrical and Computer Engineering (ECE) Department at New Jersey Institute of Technology (NJIT), where he was affiliated with the Center for Wireless Information Processing (CWiP). His research interests include information theory, machine learning, wireless communications, and neuromorphic computing. Dr Simeone is a co-recipient of the 2019 IEEE Communication Society Best Tutorial Paper Award, the 2018 IEEE Signal Processing Best Paper Award, the 2017 JCN Best Paper Award, the 2015 IEEE Communication Society Best Tutorial Paper Award and of the Best Paper Awards of IEEE SPAWC 2007 and IEEE WRECOM 2007. He was awarded a Consolidator grant by the European Research Council (ERC) in 2016. His research has been supported by the U.S. NSF, the ERC, the Vienna Science and Technology Fund, as well as by a number of industrial collaborations. He currently serves in the editorial board of the IEEE Signal Processing Magazine and is the vice-chair of the Signal Processing for Communications and Networking Technical Committee of the IEEE Signal Processing Society. He was a Distinguished Lecturer of the IEEE Information Theory Society in 2017 and 2018. Dr Simeone is a co-author of two monographs, two edited books published by Cambridge University Press, and more than one hundred research journal papers. He is a Fellow of the IET and of the IEEE. }
\end{IEEEbiography}

\begin{IEEEbiography}[{\includegraphics[width=1in,height=1.25in,clip,keepaspectratio]{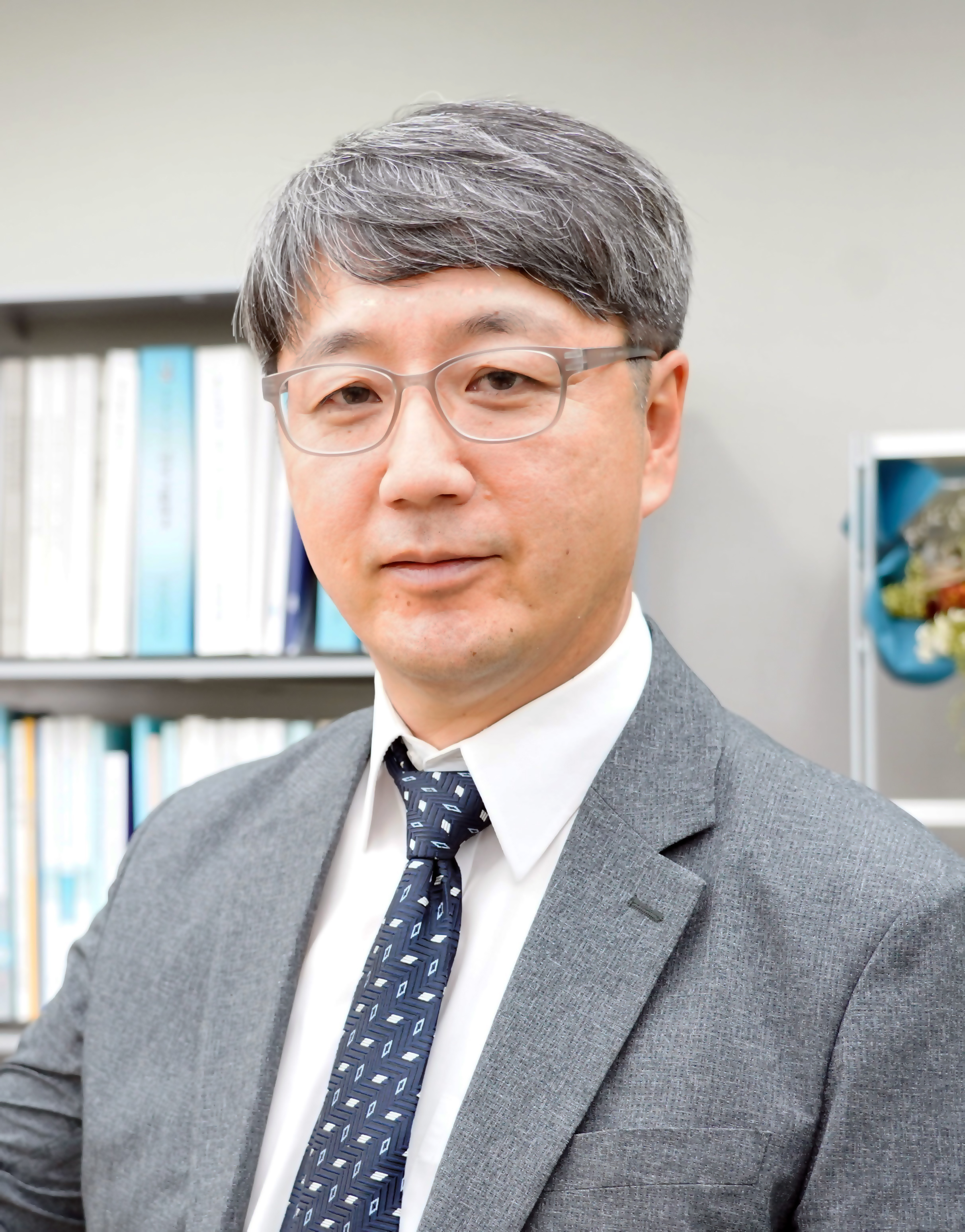}}]{Joonhyuk Kang} \textcolor{black}{(Member, IEEE) received the B.S.E. and M.S.E. degrees from Seoul National University, Seoul, South Korea, in 1991 and 1993, respectively, and the Ph.D. degree in electrical and computer engineering from The University of Texas at Austin, Austin, in 2002. From 1993 to 1998, he was a Research Staff Member at Sam- sung Electronics, Suwon, South Korea, where he was involved in the development of DSP-based real-time control systems. In 2000, he was with Cwill Telecommunications, Austin, TX, USA, where he participated in the project for multicarrier CDMA systems with antenna array. He was a Visiting Scholar with the School of Engineering and Applied Sciences, Harvard University, Cambridge, MA, USA, from 2008 to 2009. He is currently serving as Head of the School of Electrical Engineering (EE), KAIST, Daejeon, South Korea. His research interests include signal processing for cognitive radio, cooperative communication, physical-layer security, and wireless localization. He is a member of the Korea Information and Communications Society and the Tau Beta Pi (the Engineering Honor Society). He was a recipient of the Texas Telecommunication Consortium Graduate Fellowship, from 2000 to 2002.}
\end{IEEEbiography}

\end{document}